\documentclass[journal]{IEEEtran}
\usepackage{graphicx}
\usepackage[normalem]{ulem}

\usepackage[swedish,english]{babel}

\usepackage{tikz,amsmath,amssymb,amsfonts,bm}
\usepackage{pgfplots}
\newcommand{\U}{U}
\newcommand{\G}{{G}}
\newcommand{\Gp}{{G}_{\rm p}}
\newcommand{\Ep}{\vec{E}_{\rm p}}
\newcommand{\Hp}{\vec{H}_{\rm p}}
\newcommand{\Ap}{\vec{A}_{\rm p}}
\newcommand{\phip}{{\phi}_{\rm p}}
\newcommand{\JS}{\vec{J}_{\rm S}}
\newcommand{\vr}{\vec{r}}

\renewcommand{\vec}[1]{{\boldsymbol#1}}
\newcommand{\diff}{\mathrm{d}}
\providecommand*{\mrm}[1]{\mathrm{#1}}
\providecommand*{\vhat}[1]{\hat{\vec{#1}}}
\providecommand*{\iu}{\ensuremath{\mrm{j}}}
\providecommand*{\eu}{\ensuremath{\mrm{e}}}
\renewcommand{\Im}{\mathop{\mathrm{Im}}}
\renewcommand{\Re}{\mathop{\mathrm{Re}}}
\newcommand{\Om}{\Omega}
\newcommand{\ie}{\textit{i.e.}\/ }
\newcommand{\eg}{\textit{e.g.}\/ }

\newcommand{\R}{\mathbb R}

\newcommand{\Z}{\mathbb Z}

\newcommand{\We}{W_{\rm e}}
\newcommand{\Wm}{W_{\rm m}}
\newcommand{\Kzmn}{k_{{\rm z}mn}}
\newcommand{\Kzpq}{k_{{\rm z}pq}}
\newcommand{\Ktmn}{\vec{k}_{{\rm t}mn}}
\newcommand{\Ktpq}{\vec{k}_{{\rm t}pq}}

\newcommand{\Ktnull}{\vec{k}_{{\rm t}00}}
\renewcommand{\Pr}{P_{\rm r}}

\DeclareMathAlphabet\mathbfcal{OMS}{cmsy}{b}{n}
\usepackage{tikz-3dplot}
\usetikzlibrary{decorations}
\usetikzlibrary{decorations.markings}

\begin{document}

\title{Stored energies and ${\rm Q}$-factor of two-dimensionally \\ periodic antenna arrays}

\date{\vspace{-5ex}}

\author{Andrei Ludvig-Osipov* and B. L. G. Jonsson \\
{\em KTH Royal Institute of Technology,
School of Electrical Engineering}\\
{\em  and Computer Science, 100 44 Stockholm, Sweden}\\
osipov@kth.se}

\maketitle

\begin{abstract}
The Q-factor for lossless three-dimensional structures with two-dimensional periodicity is here derived in terms of the electric current density. The derivation in itself is shape-independent and based on the periodic free-space Green's function. The expression for Q-factor takes into account the exact shape of a periodic element, and permits beam steering. The stored energies and the radiated power, both required to evaluate Q-factor, are coordinate independent and expressed in a similar manner to the periodic Electric Field Integral equation, and can thus be rapidly calculated.
Numerical investigations, performed for several antenna arrays, indicate fine agreement, accurate enough to be predictive, between the proposed Q-factor and the tuned fractional bandwidth, when the arrays are not too wideband (\ie when $Q\geq 5$).
For completeness, the input-impedance Q-factor, proposed by Yaghjian and Best in 2005, is included and agrees well numerically with the derived Q-factor expression.
The main advantage of the proposed representation is its explicit connection to the current density, which allows the Q-factor to give bandwidth estimates based on the shape and current of the array element.
\end{abstract}

\section{Introduction}

The Q-factor of an oscillating system is an indirect measure of the width of the system's resonance.
In many cases, the Q-factor accurately predicts the bandwidth at which an external excitation can be efficiently applied to the system~\cite{Yaghjian+Best2005,Volakis+etal2010}.
This practically important property has encouraged a development of Q-factor representations for electric circuits~\cite{Collin1992}, electrically small antennas~\cite{Wheeler1947,Chu1948,Collin+Rothschild1964,Harrington1960,Foltz+McLean1999,Sten+etal2001}, and finite radiating structures~\cite{Vandenbosch2010,Gustafsson+Jonsson2015stored,Jonsson+Gustafsson2015}.
For periodic structures, the existing Q-factor expressions are restricted to a few geometries~\cite{Tomasic+Steyskal2007,Tomasic+Steyskal2007b,Kwon+Pozar2014}.
This paper proposes the Q-factor formulation for periodic structures with arbitrary element shape.

The Q-factor (also called quality factor, $Q$, radiation Q, antenna Q~\cite{Yaghjian+Best2005,Foltz+McLean1999,Collin+Rothschild1964}) is defined to be proportional to the ratio between cycle means of the stored and the loss energies~\cite{IEEEstandard2014}.
In electromagnetics, the relation to energies links the Q-factor with distributed quantities of the system, such as radiation modes and electric current density.
This relation paved way to many powerful Q-factor-based results in antenna theory.

For small antennas, the fundamental bounds, relating size, bandwidth and efficiency, are important examples of Q-factor results.
An essential step in establishing such bounds is the development of a suitable representation for the Q-factor.
Chu~\cite{Chu1948}, Wheeler~\cite{Wheeler1947} and Harrington~\cite{Harrington1960} obtained classical lower bounds on the Q-factor for canonical geometries by analyzing Q-factor representation, based on circuit models for spherical and cylindrical wave expansions of radiating fields.
Collin and Rothschild~\cite{Collin+Rothschild1964} used the Q-factor expressed directly in reactive (stored) and radiated energies to establish bounds for spherical and cylindrical volumes.
The current density representation of the Q-factor, proposed by Vandenbosch~\cite{Vandenbosch2010} (see also the earlier work of Geyi~\cite{Geyi2003})
and later refined by Gustafsson and Jonsson~\cite{Gustafsson+Jonsson2015stored,Jonsson+Gustafsson2015}, is the key element to establish $Q$-bounds for desired geometries using optimization methods~\cite{Gustafsson+etal2012,Vandenbosch2011,Hansen+Collin2009,Cismasu+Gustafsson2014,Tayli+etal2018,Shi+etal2017,Gustafsson+Nordebo2013}. 
A practically convenient Q-factor formula, proposed by Yaghjian and Best  for antennas with a defined port, is expressed in terms of input impedance and its derivative~\cite{Yaghjian+Best2005}, and has been used for a genetic-algorithm-based optimization of an antenna structure~\cite{Tayli+etal2018}.

For larger radiating arrays, which can be accurately described by a unit cell representation, much fewer results exist.
Tomasic and Steyskal~\cite{Tomasic+Steyskal2007,Tomasic+Steyskal2007b}  proposed a lower Q-factor bound for a 1D-array of infinitely long cylinders with either magnetic or electric current sources in free space and over a ground plane.
Kwon and Pozar~\cite{Kwon+Pozar2014} proposed a Q-factor for a 2D-flat infinite array of strip dipoles in a plane located in free space, over a ground plane, and over a dielectric substrate with a ground plane.
The key element of their derivation is a Fourier transform of the surface current density on the dipoles; the current density is there assumed to be in one direction only, and a single propagating mode is permitted.
Kwon and Pozar also suggest, that their method can be applied to other flat geometries with one-dimensional currents, however, no extension  of this idea has been published.

In this paper, we propose and derive  the Q-factor expression for three-dimensional radiating arrays with two-dimensional periodicity.
The proposed expression is a step towards developing fundamental bounds, obtainable by current density optimization, similarly to non-periodic case~\cite{Gustafsson+etal2012,Vandenbosch2011,Hansen+Collin2009,Cismasu+Gustafsson2014,Tayli+etal2018,Shi+etal2017,Gustafsson+Nordebo2013}.
The expression takes into account the exact shape of the radiating elements, and is formulated in terms of electric current density in the array's unit cell.
There is no restriction on the number of propagating Floquet modes that are permitted, and beam steering is allowed.
The current density is in general three-dimensional with no restrictions on directions of the currents.
The expression involves integration over a finite volume of current density supported within a unit cell, and can thus be rapidly calculated numerically.
The Q-factor is calculated as a function of frequency for several antenna arrays.
To validate the proposed Q-factor expression, we compare our numerical results with the tuned bandwidth, determined from data computed by a commercial full-wave solver, and also with the input-impedance-$Q$ formula of Yaghjian and Best~\cite{Yaghjian+Best2005}.
Note that our expression, in distinction to the input-impedance-$Q$, does not require a specified input port.
Additionally our expression generalizes the result of Kwon and Pozar~\cite{Kwon+Pozar2014}, which we demonstrate by restricting our result by their assumptions and geometry.

The paper is organized as follows.
Section~\ref{sec:problem_formulation} {defines} the 
periodic array {geometries}, considered in this paper, and introduces the stored energies and the radiated power for periodic structures. 
The periodic Q-factor expression, which is the main theoretical result of this paper, is derived in Sections~\ref{sec:potential_representation}--\ref{sec:source_representation}: 
in Section~\ref{sec:potential_representation}, we represent the stored energies {in terms of} electric scalar and magnetic vector potentials. {This}  is convenient for the subsequent derivation of the stored energies in terms of current density in Section~\ref{sec:source_representation}.
The complex and the radiated powers in terms of the current density are 
included for completeness in the end of Section~\ref{sec:source_representation}.
Section~\ref{sec:dipole_array} contains comparison of the proposed Q-factor with the result of Kwon and Pozar~\cite{Kwon+Pozar2014} for an array of dipoles.
In Section~\ref{sec:matrix_formulation} 
we approximate the currents by a finite basis, in order to obtain explicit matrices to facilitate  numerical calculations of the Q-factor.
Numerical examples for {a few} unit cell {antenna-}designs are presented in Section~\ref{sec:numerical}.
{We end the paper with conclusions in }Section~\ref{sec:conclusions}.

\section{Periodic arrays and stored energies}
\label{sec:problem_formulation}
In this section, we define the family of periodic arrays, that are considered in this paper.
The Q-factor, electric and magnetic stored energies, and radiated power are defined in a periodic setting.
The periodic Green's function is stated together with its associated quantities.

\begin{figure}
\centering
\includegraphics[]{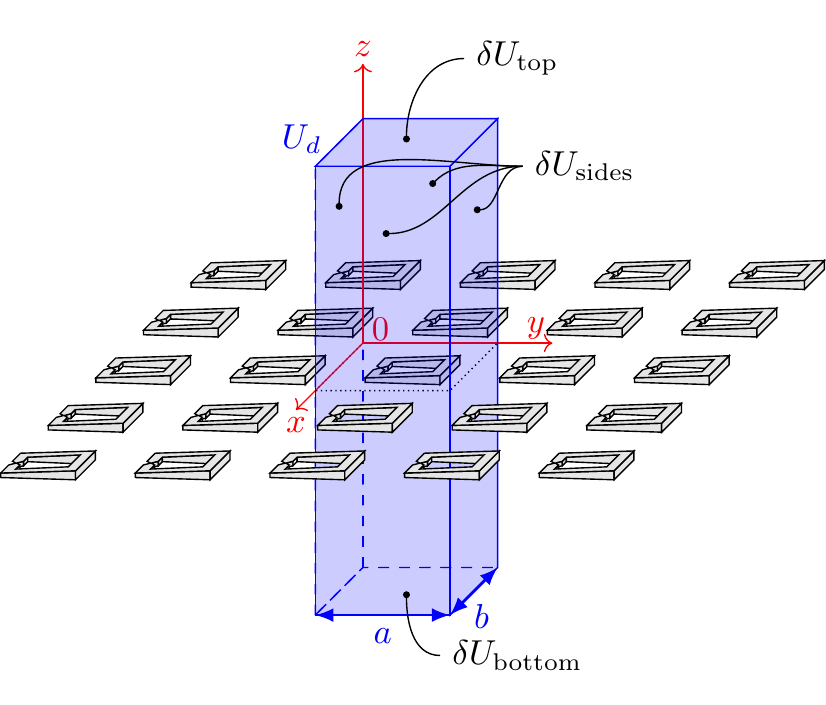}
\caption{An example of a periodic structure.}
\label{fig:UC}
\end{figure}

We consider a three-dimensional array of lossless perfectly-electric-conducting (PEC) elements of arbitrary geometry.
The array is two-dimensionally periodic on a rectangular grid. The array elements are of finite extent and sufficiently regular to support a solution to Maxwell's equations.
The unit cell is enclosed in a box $U_d=\{ \vec{r}=(x,y,z)\in\mathbb{R}^3:x\in[0,a],y\in[0,b],z\in[-d/2,d/2], a>0, b>0, d>0\}$, with the box surface denoted as $\delta \U_d$ and composed of the top surface $\delta \U_{\rm top}$ at $z=d/2$, the bottom surface $\delta \U_{\rm bottom}$ at $z=-d/2$, and the side walls $\delta \U_{\rm sides}$.
Fig.~\ref{fig:UC} shows an example of such a structure.
Here the PEC regions are shown in gray, and a unit cell $U_d$ is shown as a blue column.

The array elements support an electric current density $\vec{J}$, which is periodic up to a phase shift:
\begin{equation}
\vec{J}(\vec{r}+\vec{\zeta}_{mn})=\vec{J}(\vec{r})\eu^{\iu \Ktnull\cdot \vec{\zeta}_{mn}}. 
\label{eq:J_periodic}
\end{equation}
Here $\vec{r}\in\mathbb{R}^3$, $\vec{\zeta}_{mn}=am\hat{\vec{x}}+bn\hat{\vec{y}}$; $m,n\in \Z$, and $\Ktnull$ is a coordinate-independent transverse phasing wave-vector. The equivalent magnetic sources are neglected in this paper. The inclusion of magnetic sources into consideration can be done similarly to the finite case, see e.g.~\cite{Jonsson+Gustafsson2015}.
The goal is to find an expression for the Q-factor in terms of the unit cell electric current density~$\vec{J}$.
The time convention $\eu^{\iu \omega t}$ is used but suppressed throughout the paper, and $\omega$ and $t$ denote the angular frequency and time respectively.

The  Q factor in terms of stored electric $\We$ and magnetic $\Wm$ energies per unit cell and the total power $P_{\rm tot}$ (the radiated power and ohmic heating) per unit cell is commonly defined~\cite{IEEEstandard2014,Schab+etal2018} as
\begin{equation}
Q = \frac{2\omega \max \{ \We,\Wm\} }{P_\mrm{tot}}.
\label{eq:Q_We_Wm}
\end{equation}
For the lossless case the total power is equal to the radiated power $\Pr$.

We define the stored energies per unit cell as a difference between the total electric (magnetic) energy and radiated electric (magnetic) energy in each unit cell
\begin{equation}
\We = \frac{\epsilon}{4}
\int_{\U}|\vec{E}|^2-|\Ep|^2\diff v,
\label{eq:WeF}
\end{equation}
\begin{equation}
\Wm = \frac{\mu}{4}
\int_{\U}|\vec{H}|^2-|\Hp|^2\diff v,
\label{eq:WmF}
\end{equation}
where $\U$ is the box $\U_d$ as $d\to\infty$, $\Ep$ is the component of the $\vec{E}$-field, which includes the electric field of propagating Floquet modes only, $\vec{H}$ is the magnetic field, and $\Hp$ is the corresponding propagation modes component of the $\vec{H}$-field.
Here we note that the radiated energy is contained in the propagating Floquet modes, as they do not decay at $z\to\infty$.
This definition resembles the far-field subtraction approach for single finite-sized radiating elements~\cite{Gustafsson+Jonsson2015stored,Jonsson+Gustafsson2015}, and for a single propagating mode it is similar to the stored energies definition in~\cite{Kwon+Pozar2014}.

A convenient representation of the radiated power $\Pr$ can be obtained from the differential form of Poynting's theorem for harmonic electromagnetic fields, which is~\cite{Jackson1999}
\begin{equation}
-\nabla\cdot (\vec{E}\times \vec{H}^*) + \iu \omega (\epsilon |\vec{E}|^2-\mu |\vec{H}|^2) = \vec{E}\cdot \vec{J}^*,
\label{eq:Poyntings_theorem}
\end{equation}
where $\vec{E}$, $\vec{H}$ is the total electric and magnetic fields respectively, and the superscript $*$ indicates the complex conjugate. {Here $\epsilon$ is the free-space permittivity, and $\mu$ is the free-space permeability.}
The real part of the volume integral of~\eqref{eq:Poyntings_theorem} over the unit cell provides the relation
\begin{equation}
\begin{split}
\frac{1}{2}\Re \int_{\delta\U_d} &\left[ \vec{E}(\vec{r})\times  \vec{H}^*(\vec{r})\right]
\cdot \hat{\vec{n}}(\vec{r}) \diff S \\ &=
-\frac{1}{2}\Re \int_\Om \vec{E}(\vec{r})\cdot\vec{J}^*(\vec{r}) \diff v,
\label{eq:real_poynting}
\end{split}
\end{equation}
where $\hat{\vec{n}}$ is the outward-normal unit vector of the {boundary} surface $\delta \U_d$ of the unit cell, $\Om$ is the region of the metal inclusions within $\U_d$.
The right-hand side of~\eqref{eq:real_poynting} can be identified as the radiated power, and thus
\begin{equation}
\Pr = \Re \left\{ -\frac{1}{2} \int_\Om \vec{E}\cdot\vec{J}^* \diff v \right\} = \Re P_{\rm c},
\label{eq:P_rad}
\end{equation}
where we denoted the complex quantity in brackets as the complex power $P_{\rm c}$.

To express stored energies explicitly in terms of the unit-cell current density, we utilize the free-space Green's function for Helmholtz equation. Here we include the periodic up to a phase-shift requirement~\eqref{eq:J_periodic} on solutions from Floquet theorem by a phase correction of the point source in the unit cell:
\begin{equation}
(\nabla^2 + k^2) \G(\vec{r}_1,\vec{r}_2) = -\hspace{-1em}
\underset{(m,n)\in \Z^2}{\sum } \delta(\vec{r}_1-\vec{r}_2-\vec{\zeta_{mn}})
\eu^{-\iu \Ktnull\cdot \vec{\zeta}_{mn}}.
\end{equation}
Here $\delta(\vec{r})$ is the Dirac delta-function, $k=\omega/c_0$ is the free-space wavenumber; $c_0=1/\sqrt[]{\epsilon\mu}$ is the speed of light in free space.
The solution of this equation is the 2D-periodic Green's function.
Throughout this paper, the spectral form of the 2D-periodic Green's function is used~\cite{Oroskar+etal2006} 
\begin{equation}
\G(\vec{r}_1,\vec{r}_2)
=\frac{1}{2\iu S} \hspace{-0.5em} 
\underset{(m,n)\in \Z^2}{\sum}
\frac{1}{\Kzmn}\eu^{-\iu\Ktmn\cdot(\vec{\rho}_1-\vec{\rho}_2)}\eu^{-\iu \Kzmn|z_1-z_2|},
\label{eq:Greens_function}
\end{equation}
where $\Ktmn=\Ktnull+2\pi\frac{n}{a}\hat{\vec{x}}+2\pi\frac{m}{b}\hat{\vec{y}}$, $S=ab$, $\Kzmn=\sqrt[]{k^2-\Ktmn\cdot \Ktmn}$, and the coordinates are separated into transverse and longitudinal components $\vec{r}_i=\vec{\rho}_i+\hat{\vec{z}}z_i$, $i=\{1,2\}$. 
The transverse phasing wave-vector is $\Ktnull=k\sin\theta_0\cos\phi_0\hat{\vec{x}}+k\sin\theta_0\sin\phi_0\hat{\vec{y}}$ with the polar $\theta_0$ and azimuthal $\phi_0$ angles, which define a direction of the fundamental-mode plane wave. 
Whenever $(k^2-\Ktmn\cdot \Ktmn)$ is negative, we choose the negative branch of the square root to evaluate $\Kzmn$, so that $\eu^{-\iu \Kzmn|z_1-z_2|}$ is exponentially damped.

\section{Stored energies in terms of potentials}
\label{sec:potential_representation}
In this section we derive a potential representation of the stored energies. To obtain the current-density representation of the stored energies, we first express the electric and magnetic stored energies in terms of the electric scalar potential $\phi$ and the magnetic vector potential $\vec{A}$.
A similar potential representation approach was used in~\cite{Gustafsson+Jonsson2015stored,Jonsson+Gustafsson2015} to derive the refined expressions for the stored energies in terms of sources for single antennas.

The potentials are defined by the choice of the Lorenz gauge~\cite{Jackson1999,Landau+etal1984,vanBladel2007}
\begin{equation}
\nabla\cdot\vec{A}+\iu\omega\epsilon\mu\phi=0.
\label{eq:Lorenz}
\end{equation}
The choice of gauge implies that $(\nabla^2+k^2)\phi(\vec{r}_1)=-\varrho(\vec{r}_1)/\epsilon_0$ and $(\nabla^2+k^2)\vec{A}(\vec{r}_1)=-\mu_0\vec{J}(\vec{r}_1)$.
Here $\varrho$ is the electric charge density.
We also recall the connection between the electric field and potentials
\begin{equation}
\vec{E}=-\nabla \phi - \iu \omega \vec{A}.
\label{eq:E_phi_A}
\end{equation}

The solution to the Helmholtz equation yields the potentials expressed in terms of the sources
\begin{equation}
\phi(\vec{r}_1)=\frac{1}{\epsilon}\int_\Om \varrho(\vec{r}_2)\G(\vec{r}_1,\vec{r}_2)\diff v_2,
\label{eq:phi_G}
\end{equation}
\begin{equation}
\vec{A}(\vec{r}_1)=\mu\int_\Om \vec{J}(\vec{r}_2)\G(\vec{r}_1,\vec{r}_2)\diff v_2.
\label{eq:A_G}
\end{equation}
Note that $\varrho, \vec{J}$ have support in $\Om$, which is of finite size.
Here, the 2D-periodic Green's function $\G(\vec{r}_1,\vec{r}_2)$ is given in~\eqref{eq:Greens_function}. We observe that periodicity up to a phase-shift carries over also to the charge density, due to the continuity equation, \ie
$\varrho(\vec{r})=\varrho(\vec{r}+\vec{\zeta}_{mn})\eu^{\iu \Ktnull\cdot \vec{\zeta}_{mn}}$.

The first term of the integrand of~\eqref{eq:WeF}, the total electric energy density, is expressed using~\eqref{eq:Lorenz},~\eqref{eq:E_phi_A} and the vector identity $\vec{A}\cdot\nabla\phi^*=\nabla\cdot(\vec{A}\phi^*)-\phi^*\nabla\cdot\vec{A}$:
\begin{equation}
\begin{split}
|\vec{E}(\vec{r})|^2 
=&|\nabla\phi |^2 + \omega^2|\vec{A}|^2\\
&+2\Re\{\iu\omega\nabla\cdot(\vec{A}\phi^*)\}-2k^2|\phi|^2.
\end{split}
\label{eq:E2}
\end{equation}
Similar treatment of the electric energy density of the propagating modes gives
\begin{equation}
\begin{split}
|\Ep(\vec{r})|^2 =&
|\nabla\phip |^2 + \omega^2|\Ap|^2\\
&+2\Re\{\iu\omega\nabla\cdot(\Ap\phip^*)\}-2k^2|\phip|^2.
\label{eq:Ef2}
\end{split}
\end{equation}
The potentials $\Ap, \phip$ correspond to the propagating modes, {and} are defined as follows
\begin{equation}
\begin{split}
\phip(\vec{r}_1)=\frac{1}{\epsilon}\int_\Om \varrho(\vec{r}_2)\Gp(\vec{r}_1,\vec{r}_2)\diff v_2,
\\
\Ap(\vec{r}_1)=\mu\int_\Om \vec{J}(\vec{r}_2)\Gp(\vec{r}_1,\vec{r}_2)\diff v_2,
\label{eq:A_phi_GF}
\end{split}
\end{equation}
where the propagating-modes part of the periodic Green's function is given as
\begin{equation}
\Gp(\vec{r}_1,\vec{r}_2)
=\frac{1}{2\iu S} \hspace{-0.5em}
\underset{(m,n)\in \mathcal{P}}{\sum}
\frac{1}{\Kzmn}\eu^{-\iu\Ktmn\cdot(\vec{\rho}_1-\vec{\rho}_2)}\eu^{-\iu \Kzmn|z_1-z_2|}.
\label{eq:Greens_function_F}
\end{equation}
{Here the} summation is performed over the set of propagating modes $\mathcal{P}=\{ (m,n): k^2-\Ktmn\cdot \Ktmn \geq 0 \}$, \ie the modes for which $\Kzmn$ is purely real.

The substitution of electric energy densities~\eqref{eq:E2} and~\eqref{eq:Ef2} into the stored energy definition~\eqref{eq:WeF} gives the representation of the electric stored energy in terms of potentials
\begin{equation}
\begin{split}
&\We=\frac{\epsilon}{4}\int_\U (|\nabla\phi|^2-k^2|\phi|^2)+ \omega^2(|\vec{A}|^2-|\Ap|^2) \\ 
&- k^2(|\phi|^2-|\phip|^2) +
2\Re\{\iu\omega\nabla\cdot(\vec{A}\phi^*-\Ap\phip^*)\}\diff v.
\end{split}
\label{eq:WeF_potentials_}
\end{equation}
Here the terms are grouped for their further convenient evaluation. 
The integral of divergence term in this expression vanishes. To show that, we apply the divergence theorem~\cite{Jackson1999,Landau+etal1984,vanBladel2007}
\begin{equation}
\begin{split}
\int_\U 2\Re& \{\iu \omega \nabla\cdot (\vec{A}\phi^*-\Ap\phip^*)\} \diff v
\\
&=2\Re \{\iu\omega\int_{\delta\U}\hat{\vec{n}}\cdot(\vec{A}\phi^*-\Ap\phip^*)\diff S\}.
\end{split}
\label{eq:div_theorem_A_phi}
\end{equation}
A direct substitution of the Green's function~\eqref{eq:Greens_function} and its propagating-modes part~\eqref{eq:Greens_function_F} in potentials~\eqref{eq:phi_G},~\eqref{eq:A_G},~\eqref{eq:A_phi_GF} asserts that $\{\vec{A}(\vec{r}), \phi(\vec{r}), \Ap(\vec{r}), \phip(\vec{r}) \} \eu^{\iu \vec{k}_{\rm t 00}\cdot \vec{r}}$ are periodic with lattice vectors $a\hat{\vec{x}}$ and $b\hat{\vec{y}}$.
Evidently, $(\vec{A}\phi^*-\Ap\phip^*)$ is also periodic, and the integral over $\delta\U_{\rm sides}$ in the right-hand side of~\eqref{eq:div_theorem_A_phi} vanishes as the outward normal vectors $\vhat{n}$ are directed oppositely at the opposite side-walls.
Integration over $\delta\U_{\rm top}$ and $\delta\U_{\rm bottom}$ vanishes as $d\to\infty$, since each term of $(\vec{A}\phi^*-\Ap\phip^*)$  contains at least one evanescent-mode factor, exponentially decaying with growth of $d$.
Thus, the electric stored energy in terms of potentials is
\begin{equation}
\begin{split}
\We=\frac{\epsilon}{4}&\int_\U (|\nabla\phi|^2-k^2|\phi|^2)
\\&+ \omega^2(|\vec{A}|^2-|\Ap|^2) - k^2(|\phi|^2-|\phip|^2)\diff v.
\end{split}
\label{eq:WeF_potentials}
\end{equation}

A similar approach as above can be used to find the potential representation of the magnetic stored energy~\eqref{eq:WmF}.
Here, however, we instead integrate the expression~\eqref{eq:Poyntings_theorem} of Poynting's theorem over the unit cell volume and take the imaginary part to obtain the following relation for the imaginary part of the complex power $P_{\rm c}$
\begin{equation}
\begin{split}
\frac{\omega}{4}\int_\U& \mu|\vec{H}|^2-\epsilon|\vec{E}|^2\diff v\\
&=
-\frac{1}{4}\Im\int_\Om \vec{E}(\vec{r})\cdot\vec{J}^*(\vec{r}) \diff v = \frac{1}{2}\Im P_{\rm c},
\label{eq:imag_poynting}
\end{split}
\end{equation}
where the imaginary part of the integral of $\nabla\cdot (\vec{E}\times \vec{H}^*)$ is vanishing.
Indeed, application of the divergence theorem turns the volume integral into surface integral of $\vec{E}\times \vec{H}^*$ over the surface of $\U$. Due to the periodicity of $\vec{E}\times \vec{H}^*$, integral over side walls $\delta\U_{\rm sides}$ vanishes. The integral over unit cell's top $\delta \U_{\rm top}$ and bottom $\delta \U_{\rm bottom}$ surfaces is purely real, which is shown in Appendix~\ref{app:vanishingImExH}.
Combining~\eqref{eq:imag_poynting} with the plane-wave behavior $\int_{\U_d} (|\Hp|^2-|\Ep|^2/\eta^2) \diff v\rightarrow 0$ as $d\to\infty$
and the definitions of the stored energies~\eqref{eq:WeF} and~\eqref{eq:WmF}, we obtain the connection between the electric and magnetic energies
\begin{equation}
\Wm = \We + \frac{1}{2\omega} \Im P_{\rm c}.
\label{eq:Wm_Pc}
\end{equation}
Here, $\eta=\sqrt[]{\mu/\epsilon}$
is the free space impedance for the propagating modes.
The complex power, $P_{\rm c}$, is formulated in terms of sources in the end of Section~\ref{sec:source_representation}.

\section{Stored energies in terms of sources}
\label{sec:source_representation}
In the potential formulation of the electric stored energy~\eqref{eq:WeF_potentials}, we have combined the integrand terms in pairs. 
The pairs are chosen such that the integral of each pair is finite.
In this section, we derive from these pairs the representation of the stored energy in terms of the electric current density $\vec{J}$.

We begin with the evaluation of the integrand's first pair, $(|\nabla\phi|^2-k^2|\phi|^2)$, which is a weak form of a Helmholtz equation. Using the vector identity $|\nabla\phi|^2=\nabla\cdot(\phi\nabla\phi^*)-\phi\Delta\phi^*$ and the Helmholtz equation for the scalar potential, we obtain that
\begin{equation}
\begin{split}
|\nabla\phi|^2-k^2|\phi|^2 =
\Re \{ \nabla\cdot(\phi\nabla\phi^*) + \phi\varrho^*/\epsilon\}.
\end{split}
\end{equation}
{The divergence term in this expression vanish after an integration over $\U$, to see this note that the divergence theorem gives}
\begin{equation}
\begin{split}
\int_\U\Re \{ \nabla\cdot(\phi\nabla\phi^*)\}\diff v =
\int_{\delta\U}\Re \{\hat{\vec{n}}\cdot (\phi\nabla\phi^*)\}\diff S,
\end{split}
\end{equation}
where $\delta\U$ is the boundary of $\U$.
Periodicity of $(\phi\nabla\phi^*)$ follows from periodicity of $\phi(\vec{r}) \eu^{\iu \vec{k}_{\rm t 00}\cdot \vec{r}}$. 
This, in turn, means that the integrals over the opposite side walls 
of $\U$ cancel each other. Additionally, the integral over the top and the bottom of $\U$ vanishes, since $(\phi\nabla\phi^*)$ is purely imaginary {for} large $|z|$. 
{For the details of this derivation, see} Appendix~\ref{app:term1_vanishing}.
Thus we have
\begin{equation}
\begin{split}
\int_\U|\nabla\phi|^2-k^2|\phi|^2\diff v =
\frac{1}{\epsilon}\Re \{ \int_\Om \phi\varrho^*\diff v\}.
\end{split}
\label{eq:phi_rho}
\end{equation}
A combination of the Green's function representation of the scalar potential~\eqref{eq:phi_G} and the continuity equation 
\begin{equation}
\iu\omega\varrho=-\nabla\cdot\vec{J}
\label{eq:continuity}
\end{equation}
for current density yields the expression for the scalar potential in terms of currents
\begin{equation}
\phi(\vec{r}_1)=
\frac{\iu}{\omega\epsilon}\int_\Om \G(\vec{r}_1,\vec{r}_2)\nabla_2\cdot\vec{J}(\vec{r}_2) \diff v_2.
\label{eq:phi_J}
\end{equation}
The continuity equation~\eqref{eq:continuity} gives us the electric current representation of the charge density $\varrho$.
Substitution of $\varrho$ and $\phi$ into~\eqref{eq:phi_rho} yields
the source representation of the first pair as a quadratic form
\begin{equation}
\begin{split}
&W_{\rm e,1} = \frac{\epsilon}{4}
\int_\U|\nabla\phi|^2-k^2|\phi|^2\diff v \\ &=
 \frac{\mu}{4k^2}\Re 
\int_\Om \int_\Om (\nabla_1\cdot \vec{J}^*(\vec{r}_1)) \G(\vec{r}_1,\vec{r}_2) \nabla_2\cdot \vec{J}(\vec{r}_2) \diff v_2 \diff v_1
 \\&= \langle \vec{J},\mathcal{W}_{\rm e,1}\vec{J}\rangle,
\end{split}
\raisetag{\normalbaselineskip}
\label{eq:We1}
\end{equation}
where the scalar product $\langle \vec{J},C(\vec{J})\rangle=\int_{\Omega} \vec{J}^*\cdot C(\vec{J})\diff v_1$ and  
$\mathcal{W}_{\rm e,1}$ is the linear operator acting on $\vec{J}$. 
The above divergence form~\eqref{eq:We1} is equivalent to the operator $\mathcal{W}_{\rm e,1}$ acting on  the set of current densities $\vec{J}$ considered here. 

To evaluate the second pair in the stored energy representation~\eqref{eq:WeF_potentials}, we use~\eqref{eq:A_G} to express the vector potential squared magnitude
\begin{equation}
\begin{split}
|\vec{A}(\vec{r})|^2 = 
&\mu^2
\int_\Om \int_\Om \G^*(\vec{r},\vec{r}_1) \G(\vec{r},\vec{r}_2) \vec{J}^*(\vec{r}_1)\cdot\vec{J}(\vec{r}_2)\diff v_2 \diff v_1.
\end{split}
\label{eq:A_J}
\end{equation}
We can similarly express the contribution due to the vector potential of the propagating modes. We utilize~\eqref{eq:A_phi_GF} to get
\begin{equation}
\label{eq:F_A_J}
\begin{split}
|\Ap&(\vec{r})|^2 \\
=& \mu^2 \hspace{-0.5em}
\int_\Om \int_\Om
\Gp^*(\vec{r},\vec{r}_1) \Gp(\vec{r},\vec{r}_2)\vec{J}^*(\vec{r}_1)\cdot \vec{J}(\vec{r}_2)\diff v_2 \diff v_1.
\end{split}
\end{equation}
Direct substitution of~\eqref{eq:A_J} and~\eqref{eq:F_A_J} into the second pair of~\eqref{eq:WeF_potentials} together with an interchange of integration order, see Appendix~\ref{app:term2}, yields
\begin{equation}
\begin{split}
W_{\rm em,1} &=
\frac{\epsilon\omega^2}{4} \int_\U (|\vec{A}(\vec{r})|^2-|\Ap(\vec{r})|^2)\diff v \\
&=\frac{\mu k^2}{4}
\int_\Om \int_\Om g(\vec{r}_1,\vec{r}_2)
\vec{J}^*(\vec{r}_1)\cdot\vec{J}(\vec{r}_2)\diff v_1 \diff v_2 \\ 
&=\langle \vec{J},\mathcal{W}_{\rm em,1}\vec{J}\rangle, 
\end{split}
\label{eq:We2}
\end{equation}
where the kernel is
\begin{equation}
\begin{split}
 &g(\vec{r}_1,\vec{r}_2) 
 \\&= \int_\U \left\{ \G^*(\vec{r},\vec{r}_1) \G(\vec{r},\vec{r}_2)  -
\Gp^*(\vec{r},\vec{r}_1) \Gp(\vec{r},\vec{r}_2) \right\} \diff v.
\label{eq:low_case_g}
\end{split}
\end{equation}
In order to simplify $g$, we start with utilizing the structure of the integrand and note that the subtraction of sums in the respective Green's functions reduce the integrand to summation over $(m,n,p,q)\in\Z^4 \setminus \mathcal{P}^2$ of terms of the form 
\begin{equation}
\begin{split}
\frac{1}{4 S^2} 
\frac{1}{\Kzmn^*\Kzpq}
\eu^{\iu\Ktmn\cdot(\vec{\rho}-\vec{\rho}_1)} &
\eu^{-\iu\Ktpq\cdot(\vec{\rho}-\vec{\rho}_2)} \\
&\eu^{\iu \Kzmn|z-z_1|}\eu^{-\iu \Kzpq^*|z-z_2|}.
\end{split}
\end{equation}
Furthermore,  we utilize the Floquet modes orthogonality under integration in the $xy$-plane over the unit-cell~\eqref{eq:orthogonality} to obtain a sum over $\Z^2\setminus \mathcal{P}$.
The subsequent integration in $z$ reduces the kernel to
\begin{equation}
\begin{split}
g(\vec{r}_1,\vec{r}_2)  
=\frac{1}{4 S} &
\underset{(m,n)\in \Z^2\setminus \mathcal{P}}{\sum} 
\frac{1}{|\Kzmn|^2} 
\eu^{\iu \Ktmn\cdot(\vec{\rho}_1-\vec{\rho}_2)} \\&
\eu^{-|\Kzmn||z_2-z_1|} \left( \frac{1}{|\Kzmn|} + |z_1-z_2|  \right),
\label{eq:little_g_exact}
\end{split}
\end{equation}
as shown in Appendix~\ref{app:term2}.
{The rather remarkable reduction in complexity in going from~\eqref{eq:low_case_g} to \eqref{eq:little_g_exact} facilitates an effective numerical computation of $g$, in particular with the exponential decay in $m,n$ of the terms in the sum.}
From the conjugate symmetry $g(\vec{r}_1,\vec{r}_2)=g^*(\vec{r}_2,\vec{r}_1)$ it follows immediately that $W_{\rm em,1}$ is purely real.

When $z_1\neq z_2$, the convergence of the double sum in~\eqref{eq:little_g_exact} is ensured by the exponential damping of the terms with growth of both $m$ and $n$.
For the special case with flat thin antennas parallel with the periodicity plane, we have $z_1=z_2$, and the kernel reduces to
\begin{equation}
\begin{split}
& g_0(\vec{r}_1,\vec{r}_2) =
\frac{1}{4 S}
\underset{(m,n)\in \Z^2\setminus \mathcal{P}}{\sum}
\frac{1}{|\Kzmn|^3} 
\eu^{\iu \Ktmn\cdot(\vec{\rho}_1-\vec{\rho}_2)},
\end{split}
\label{eq:smallG}
\end{equation}
where the terms decay at least with the cubic power of both $m$ and $n$. 
We observe that both \eqref{eq:little_g_exact} and~\eqref{eq:smallG} are independent of the choice of the coordinate system. This is markedly different from the finite support antenna case in which the far-field subtraction results in a weakly coordinate dependent stored electric energy~\cite{Gustafsson+Jonsson2015}.

We apply {a} similar procedure to the last pair in the electric stored energy representation~\eqref{eq:WeF_potentials}, which involves the scalar potential and its propagating-mode part.
{Substitution of}~\eqref{eq:phi_J} yields that
\begin{equation}
\begin{split}
|\phi(\vec{r})|^2 =
\frac{1}{\omega^2\epsilon^2}\int_\Om\int_\Om & \G^*(\vec{r},\vec{r}_1)\G(\vec{r},\vec{r}_2)\\
& \nabla_1\cdot\vec{J}^*(\vec{r}_1) 
  \nabla_2\cdot\vec{J}(\vec{r}_2) \diff v_2\diff v_1.
\end{split}
\end{equation}
In the same manner we can find the expression for the scalar potential contribution due to propagating modes
\begin{equation}
\begin{split}
|\phip(\vec{r})|^2 = \frac{1}{\omega^2\epsilon^2}
\int_\Om \int_\Om & \Gp^*(\vec{r},\vec{r}_1) \Gp(\vec{r},\vec{r}_2) \\ 
&\nabla_1\cdot\vec{J}^*(\vec{r}_1)
 \nabla_2\cdot\vec{J}(\vec{r}_2) \diff v_2 \diff v_1.
\end{split}
\end{equation}
Combining the latter two expressions {yields} the third pair in~\eqref{eq:WeF_potentials} {which reduces to the same kernel $g$ in \eqref{eq:little_g_exact} above. Thus}
\begin{equation}
\begin{split}
W_{\rm em,2} &=
\frac{\epsilon k^2}{4} \int_\U (|\phi|^2-|\phip|^2)\diff v \\
=&\frac{\mu}{4}
\int_\Om \int_\Om g(\vec{r}_1,\vec{r}_2)
\nabla_1\cdot\vec{J}^*(\vec{r}_1)
 \nabla_2\cdot\vec{J}(\vec{r}_2)\diff v_2 \diff v_1 \\
 =&
 \langle \vec{J},\mathcal{W}_{\rm em,2}\vec{J}\rangle.
\end{split}
\label{eq:We3}
\end{equation}
From the conjugate symmetry of the kernel $g$ it follows that $W_{\rm em,2}$ is real.
Finally, we  combine~\eqref{eq:We1},~\eqref{eq:We2} and~\eqref{eq:We3} to write the current representation of the electric stored energy
\begin{equation}
\begin{split}
\We &= W_{\rm e,1} + W_{\rm em,1} - W_{\rm em,2} \\
&= \langle \vec{J},(\mathcal{W}_{\rm e,1}+\mathcal{W}_{\rm em,1}-\mathcal{W}_{\rm em,2})\vec{J}\rangle.
\end{split}
\label{eq:We_source}
\end{equation}

To find the current density representation of the radiated power~\eqref{eq:P_rad} and the magnetic stored energy via~\eqref{eq:Wm_Pc}, we need to evaluate the complex power 
\begin{equation}
P_{\rm c} = 
-\frac{1}{2}\int_\Om\vec{E}\cdot\vec{J}^*\diff v.
\end{equation}
We proceed with evaluating the integrand by substituting the potential form of the electric field~\eqref{eq:E_phi_A} and consequently the potentials in terms of currents~\eqref{eq:A_G} and~\eqref{eq:phi_J}
\begin{equation}
\begin{split}
\vec{E}&\cdot\vec{J}^*=
(-\nabla_1 \phi(\vec{r}_1) - \iu \omega \vec{A}(\vec{r}_1))\cdot \vec{J}^*(\vec{r}_1) \\
&=
\left(-\nabla_1 \frac{\iu}{\omega\epsilon}\int_\Om \G(\vec{r}_1,\vec{r}_2)\nabla_2\cdot\vec{J}(\vec{r}_2) \diff v_2 \right. \\
&\qquad\qquad - \left. \iu \omega \mu\int_\Om \vec{J}(\vec{r}_2)\G(\vec{r}_1,\vec{r}_2)\diff v_2\right)\cdot \vec{J}^*(\vec{r}_1).
\end{split}
\end{equation}
By applying the vector identity 
\begin{equation}
\begin{split}
\vec{J}^*(\vec{r}_1)\cdot\nabla_1 \G(\vec{r}_1,\vec{r}_2) 
=&
-\G(\vec{r}_1,\vec{r}_2)\nabla_1\cdot\vec{J}^*(\vec{r}_1) \\ &+ \nabla_1 \cdot \{ \vec{J}^*(\vec{r}_1)\G(\vec{r}_1,\vec{r}_2) \},
\end{split}
\end{equation}
and the fact that
\begin{equation}
    \int_\Om \nabla_1 \cdot \{ \vec{J}^*(\vec{r}_1)\G(\vec{r}_1,\vec{r}_2) \}\diff v_1=0,
\end{equation}
we rewrite the complex power as
\begin{equation}
\begin{split}
P_{\rm c} = 
\frac{\iu}{2}\int_\Om\int_\Om & 
\frac{-\eta}{k} \G(\vec{r}_1,\vec{r}_2)\nabla_1\cdot\vec{J}^*(\vec{r}_1) \nabla_2\cdot\vec{J}(\vec{r}_2) \\ &+
k\eta\G(\vec{r}_1,\vec{r}_2)\vec{J}^*(\vec{r}_1)\cdot\vec{J}(\vec{r}_2)
\diff v_1 \diff v_2\\
&=\frac{1}{2} \langle \vec{J},\mathcal{Z}\vec{J}\rangle. 
\label{eq:Pc_J}
\end{split}
\end{equation}
Note that the kernel of the operator $\mathcal{Z}$ is the same as the electric field integral equation (EFIE) kernel in the method of moments (MoM) for EM-simulations~\cite{Rao+etal1982}.
We notice that the imaginary part of the first term of~\eqref{eq:Pc_J} is $-2\omega W_{\rm e,1}$.
Equation~\eqref{eq:Wm_Pc} can thus be used to obtain the magnetic stored energy
\begin{equation}
    \Wm = W_{\rm m,1} + W_{\rm em,1} - W_{\rm em,2},
\end{equation}
where
\begin{equation}
\begin{split}
W_{\rm m,1} &= 
\frac{\mu}{4}\Re \int_\Om\int_\Om 
\G(\vec{r}_1,\vec{r}_2)\vec{J}^*(\vec{r}_1)\cdot\vec{J}(\vec{r}_2)
\diff v_1 \diff v_2 \\&=
\langle \vec{J},\mathcal{W}_{\rm m,1}\vec{J}\rangle.
\label{eq:Wm1}
\end{split}
\end{equation}
Thus
\begin{equation}
\begin{split}
\Wm =
\langle \vec{J},(\mathcal{W}_{\rm m,1}+\mathcal{W}_{\rm em,1}-\mathcal{W}_{\rm em,2})\vec{J}\rangle.
\label{eq:Wm_source}
\end{split}
\end{equation}
For the case of PEC structures, only the currents on the surfaces of the elements exist.
The adjustment of the expressions for stored energies and complex power to surface currents is straightforwardly done by replacing the volume current density $\vec{J}$ with the surface current density $\JS$ and the volume elements $\diff v$ with the surface elements $\diff S$ in the integrals in~\eqref{eq:We1}~\eqref{eq:We2},~\eqref{eq:We3},~\eqref{eq:Pc_J} and~\eqref{eq:Wm1}.

\section{Array of dipoles: comparison with Kwon and Pozar}
\label{sec:dipole_array}
In this section, we examine our proposed expressions for stored energies, radiated power and the Q-factor by considering a particular array element shape -- narrow, infinitely thin rectangular dipoles situated in the array plane.
This case was considered earlier by Kwon and Pozar~\cite{Kwon+Pozar2014} for a single propagating mode. There, they assume only $x$-directed currents on rectangular dipoles contained in a plane, parallel to the $xy$-plane:
\begin{equation}
\vec{J}_0(\vec{r}) = \hat{\vec{x}}f(\vec{r})
\label{eq:Js_kwon}
\end{equation}
with $f(\vec{r})=f(x,y)$ an electric current density profile.
Below, we illustrate that our Q-factor expression, under these assumptions on the current density and number of propagating modes, coincides with the result of~\cite{Kwon+Pozar2014}.

Using vector calculus identities, we can rephrase~\eqref{eq:We1} as
\begin{equation}
\begin{split}
W_{\rm e,1} = 
\Re \frac{-\mu}{4k^2} &\int_\Om \int_\Om  \vec{J}_0^*(\vec{r}_1) \\
&\cdot \left[ \vec{J}_0(\vec{r}_2) \cdot \overline{\overline{\nabla}}_1\nabla_1 \G(\vec{r}_1,\vec{r}_2) \right] \diff S_1 \diff S_2,
\end{split}
\end{equation}
where $\overline{\overline{\nabla}}_1\nabla_1 \G(\vec{r}_1,\vec{r}_2)$ should be interpreted as a Jacobian of $\nabla_1 \G(\vec{r}_1,\vec{r}_2)$.
The substitution of~\eqref{eq:Js_kwon} for the surface current densities yields
\begin{equation}
W_{\rm e,1} = 
\Re \frac{-\mu}{4k^2} \int_\Om \int_\Om f^*(\vec{r}_1) f(\vec{r}_2) \frac{\partial^2}{\partial x_1^2} \G(\vec{r}_1,\vec{r}_2) \diff S_1 \diff S_2.
\end{equation}
By inserting the Green's function~\eqref{eq:Greens_function} with $|z_1-z_2|=0$, and taking the second order derivative, we obtain
\begin{equation}
\begin{split}
&W_{\rm e,1} = 
\Re \frac{\mu}{8k^2ab} \underset{(m,n)\in \Z^2}{\sum }
\frac{k_{{\rm x}m}^2}{\iu \Kzmn} \int_\Om f^*(\vec{r}_1) \eu^{-\iu \Ktmn \cdot \vec{\rho}_1} \diff S_1 \\ &\int_\Om f(\vec{r}_2) \eu^{\iu \Ktmn \cdot \vec{\rho}_2}  \diff S_2= 
\Re \frac{\mu}{8k^2ab} \underset{(m,n)\in \Z^2}{\sum }
\frac{k_{{\rm x}m}^2}{\iu\Kzmn} |F_{mn}|^2,
\label{eq:We1_Fmn_intermediate}
\end{split}
\end{equation}
where $k_{{\rm x}m}=\Ktmn\cdot\hat{\vec{x}}$, and $F_{mn} = \int_\Om f(\vec{r}_1) \eu^{\iu \Ktmn \cdot \vec{\rho}} \diff S$ are the Fourier transform coefficients.
We recall that $\Kzmn\in\R$ for propagating modes, and hence the propagating-modes terms $(m,n)\in\mathcal{P}$ in the sum are purely imaginary and do not contribute to $W_{\rm e,1}$.
For the evanescent modes, $\Kzmn = -\iu |\Kzmn|$, which follows from the branch choice.
This transforms~\eqref{eq:We1_Fmn_intermediate} into
\begin{equation}
W_{\rm e,1} = 
\frac{\mu}{8k^2ab} \underset{(m,n)\in \Z^2\setminus\mathcal{P}}{\sum }
\frac{k_{{\rm x}m}^2}{|\Kzmn|} |F_{mn}|^2,
\end{equation}
where Kwon and Pozar assume~\cite{Kwon+Pozar2014} that only one mode propagates, \ie $\mathcal{P}=\{(0,0)\}$.
A treatment of~\eqref{eq:We3} with similar steps yields
\begin{equation}
\begin{split}
W_{\rm em,2} &= 
\frac{-\mu}{4k^2} \int_\Om \int_\Om \vec{J}_0^*(\vec{r}_1) \\
&\hspace{5em}\cdot \left[ \vec{J}_0(\vec{r}_2) \cdot \overline{\overline{\nabla}}_1\nabla_1 g(\vec{r}_1,\vec{r}_2) \right] \diff S_1 \diff S_2 \\ &=
\frac{\mu}{16k^2ab} \underset{(m,n)\in \Z^2\setminus\mathcal{P}}{\sum }
\frac{k_{{\rm x}m}^2k^2}{|\Kzmn|^3} |F_{mn}|^2.
\end{split}
\end{equation}
By identifying the Fourier coefficients as done in~\eqref{eq:We1_Fmn_intermediate}, we easily obtain the other two contributions~\eqref{eq:We1},~\eqref{eq:Wm1} of the stored energies
\begin{equation}
W_{\rm em,1} = \frac{\mu}{16k^2ab} \underset{(m,n)\in \Z^2\setminus\mathcal{P}}{\sum }
\frac{k^4}{|\Kzmn|^3} |F_{mn}|^2,
\end{equation}
\begin{equation}
W_{\rm m,1} =
\frac{\mu}{8ab} \underset{(m,n)\in \Z^2\setminus\mathcal{P}}{\sum }
\frac{1}{|\Kzmn|} |F_{mn}|^2.
\end{equation}
Combining the stored-energies terms according to~\eqref{eq:We_source} and~\eqref{eq:Wm_source}, we obtain
\begin{equation}
\We = 
 \frac{\mu}{16k^2ab} \hspace{-2em} \underset{(m,n)\in \Z^2\setminus\{(0,0)\}}{\sum } \left[
\frac{k^2(k^2-k_{{\rm x}m}^2)}{|\Kzmn|^2} + 2k_{{\rm x}m}^2 \right] \frac{|F_{mn}|^2}{|\Kzmn|},
\label{eq:We_kwon}
\end{equation}
\begin{equation}
\Wm = 
\frac{\mu}{16ab} \hspace{-1.5em} \underset{(m,n)\in \Z^2\setminus\{(0,0)\}}{\sum } \left[
\frac{k_{{\rm y}n}^2}{|\Kzmn|^2} + 1 \right] \frac{|F_{mn}|^2}{|\Kzmn|},
\label{eq:Wm_kwon}
\end{equation}
where $k_{{\rm y}n}=\Ktmn\cdot\hat{\vec{y}}$.

The radiated power is found similarly by combining~\eqref{eq:P_rad} and~\eqref{eq:Pc_J} with the derivation steps used in this section
\begin{equation}
\begin{split}
\Pr &= \Re \frac{\iu}{2}\int_\Om\int_\Om 
\frac{\eta}{k}\vec{J}_0^*(\vec{r}_1) \cdot \left[ \vec{J}_0(\vec{r}_2) \cdot \overline{\overline{\nabla}}_1\nabla_1 \G(\vec{r}_1,\vec{r}_2) \right] \\
& \hspace{6em}
+
k\eta\G(\vec{r}_1,\vec{r}_2)\vec{J}_0^*(\vec{r}_1)\cdot\vec{J}_0(\vec{r}_2)
\diff S_1 \diff S_2 \\ &=
\frac{\eta}{4kab} \underset{(m,n)\in \mathcal{P}}{\sum } \frac{k^2-k_{{\rm x}m}^2}{\Kzmn}|F_{mn}|^2 
\\&=
\frac{\eta(1 - \sin^2\theta_0\cos^2\phi_0)}{4ab\cos\theta_0}|F_{00}|^2.
\end{split}
\raisetag{2\normalbaselineskip}
\label{eq:Pr_kwon}
\end{equation}
The expressions~\eqref{eq:We_kwon}-\eqref{eq:Pr_kwon} are exactly coinciding with the result of Kwon and Pozar~\cite{Kwon+Pozar2014}. Note that a difference by a $1/2$ factor, caused by a difference in definitions, is both in the stored energies and the radiated power, and is eliminated when the Q-factor is computed.

\section{Matrix formulation of stored energies and radiated power}
\label{sec:matrix_formulation}
In this section we discuss how the source representations of the stored energies~\eqref{eq:We1}, \eqref{eq:We2}, \eqref{eq:We3}, \eqref{eq:Wm_source} and the complex power~\eqref{eq:Pc_J}, together with the EFIE-based MoM solver, can be used to compute the Q-factor for a 2D periodic array of an arbitrary shape.
We start with introducing a set of basis functions $\{\vec{f}_n(\vec{r})\}$, which approximates the surface current density $\JS$ 
on $\Om$
\begin{equation}
\JS(\vec{r}) \approx \sum_{n=1}^{N} I_n \vec{f}_n(\vec{r}),
\end{equation}
where ${\rm \bf I}=(I_1, I_2, ..., I_N)^{\rm T}$ is a vector of the current density coefficients, $N$ is the number of the basis functions.
We proceed with substituting this basis expansion in
the components of the electric stored energies~\eqref{eq:We1},~\eqref{eq:We2} and~\eqref{eq:We3}.
This yields the matrix representation of the  electric stored energy~\eqref{eq:We_source}
\begin{equation}
W_{\rm e} = 
\langle \JS,(\mathcal{W}_{\rm e,1}+\mathcal{W}_{\rm em,1}-\mathcal{W}_{\rm em,2})\JS\rangle \approx
{\rm {\bf I}^H {\bf W}_e {\bf I}},
\end{equation}
where the kernel matrix elements are given by ${\rm \bf W}_{\rm e}^{(m,n)}=\langle \vec{f}_m,(\mathcal{W}_{\rm e,1}+\mathcal{W}_{\rm em,1}-\mathcal{W}_{\rm em,2})\vec{f}_n\rangle$, and the superscript $(.)^{\rm H}$ denotes the Hermitian transpose.
We continue in this fashion obtaining the matrix form of the complex power
\begin{equation}
P_{\rm c} \approx \frac{1}{2}{\rm {\bf I}^H {\bf Z} {\bf I}},
\end{equation}
where the impedance matrix elements are given as ${\rm \bf Z}^{(m,n)}=\langle \vec{f}_m,\mathcal{Z}\vec{f}_n\rangle$.
In accordance with~\eqref{eq:Wm_Pc}, we find the kernel matrix for the magnetic stored energy
\begin{equation}
{\rm \bf W}_{\rm m} = {\rm \bf W}_{\rm e} + \frac{\Im {\rm  {\bf Z}}}{4\omega}.
\label{eq:Wm_matrix}
\end{equation}
Combination of the above matrix representations with~\eqref{eq:Q_We_Wm} and~\eqref{eq:P_rad} gives the matrix forms for the electric and magnetic Q-factors, and the total Q-factor respectively
\begin{equation}
\begin{gathered}
Q_{\rm e} \approx \frac{4\omega {\rm {\bf I}^H {\bf W}_e {\bf I}}}{{\rm {\bf I}^H (\Re{\bf Z}) {\bf I}}}, \quad
Q_{\rm m} \approx \frac{4\omega {\rm {\bf I}^H {\bf W}_m {\bf I}}}{{\rm {\bf I}^H (\Re{\bf Z}) {\bf I}}}, \\
Q=\max \{ Q_{\rm e},Q_{\rm m} \}.
\label{eq:Qe_Qm}
\end{gathered}
\end{equation}
To calculate the $Q$-factor for a given geometry and excitation, the current density coefficients ${\rm \bf I}$ can be computed \eg by solving the EFIE by MoM, as done in the next section.
When the goal is to find the optimal-bandwidth currents on a given spatial support, the Q-factor~\eqref{eq:Qe_Qm} can be minimized with respect to the current density coefficients ${\rm \bf I}$, see finite antenna examples in \eg~\cite{Gustafsson+etal2012,Vandenbosch2011,Gustafsson+Nordebo2013}.

\section{Numerical results}
\label{sec:numerical}
The goal of this section is to illustrate the practical use of the proposed Q-factor expression.
We numerically validate the Q-factor representation by comparison with the actual tuned-impedance  bandwidth, and the input-impedance Q-factor by Yaghjian and Best~\cite{Yaghjian+Best2005}. Additionally, we compare our Q-factor expression with the Q-factor from~\cite{Kwon+Pozar2014} for a simple dipole array. 
For the case of narrow-band arrays, we illustrate that all Q-factors give similar results
and that they predict the fractional bandwidth.

We start with recalling how the actual tuned bandwidth is obtained and converted into the equivalent-tuned-bandwidth Q-factor $Q_{\rm B}$; and also how the input-impedance Q-factor $Q_{\rm Z}$ is calculated.

The approach to find $Q_{\rm B}$ follows Yaghjian and Best~\cite{Yaghjian+Best2005}: given an input impedance $Z(\omega)$ of an unmatched antenna, we tune it at the angular frequency $\omega_0$ with a series element of impedance $\iu X_{\rm s}(\omega)$ such that $\Im Z(\omega_0)+X_{\rm s}(\omega_0)=0$. 
The series element can either be an inductance $L$ with $X_{\rm s}(\omega)=\omega L$ or a capacitor $C$ with $X_{\rm s}(\omega)=-1/(\omega C)$ depending on a sign of $\Im Z$.
The overall tuned input impedance is thus $Z_0(\omega)=Z(\omega)+\iu X_{\rm s}(\omega)$, and the characteristic impedance $Z_{\rm ch}$ is set as a constant equal to $\Re Z(\omega_0)$.
The reflection coefficient is then calculated as $\Gamma = (Z_0-Z_{\rm ch})/(Z_0+Z_{\rm ch})$ and the fractional bandwidth $B$ is estimated at the threshold level $\Gamma_0$.
Finally the fractional bandwidth is converted to an equivalent Q-factor~\cite{Yaghjian+Best2005} by
\begin{equation}
Q_{\rm B} = \frac{2\Gamma_0}{B\sqrt{1-\Gamma_0^2}}.
\end{equation}

The Yaghjian and Best formula provides an estimate of the  Q{-factor}, matched at a given angular frequency, and involves the antenna's input impedance $(R+\iu X)$
\begin{equation}
Q_{\rm Z}(\omega) = \frac{\omega}{2R(\omega)}
\sqrt[]{ [ R'( \omega ) ]^2 + 
[ X'( \omega ) + |X(\omega)|/\omega]^2},
\end{equation}
where $(.)'$ denotes the angular frequency derivative.

Below we compare the here derived Q-factor with the $Q_{\rm B}$ and $Q_{\rm Z}$, computed from an input impedance given both by our in-house MoM code and CST MW Studio simulation in the frequency domain.
The implementation of the proposed Q-factor expression is based on and is an extension of our in-house MoM code~\cite{Maldonado2011}, where RWG basis functions~\cite{Rao+etal1982} were used.
In all the examples, both CST and the in-house-MoM-code models have voltage-gap excitation. 
We used Ewald's method~\cite{Oroskar+etal2006} to efficiently evaluate the Green's function~\eqref{eq:Greens_function}, for alternative acceleration methods see \eg\cite{Valerio+etal2007}.
The convergence of our MoM results are validated by comparison of input impedances with solutions obtained from CST MW Studio 2017.

\subsection{Dipole array}
We first consider an infinite planar array of strip dipoles.
This example has been treated in Section~\ref{sec:dipole_array}, where we analytically imposed the assumptions of~\cite{Kwon+Pozar2014} by Kwon and Pozar to compare our expressions with their result.
Here, however, we do not apply the assumptions of~\cite{Kwon+Pozar2014} when calculating Q-factor with our expression~\eqref{eq:Qe_Qm}.

\begin{figure}[htbp]
\centering
  \begin{minipage}{0.4\textwidth}
  \centering
    \begingroup
	\includegraphics[]{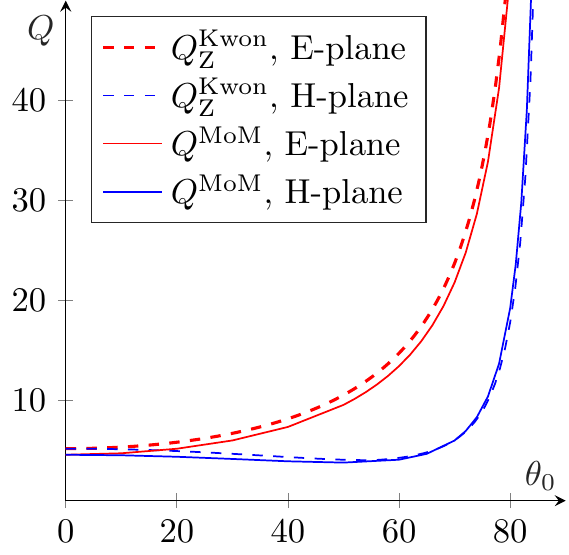}
	
	(a)
  \endgroup
  \end{minipage}
  \begin{minipage}{0.5\textwidth}
  \centering
  \begingroup
	\includegraphics[]{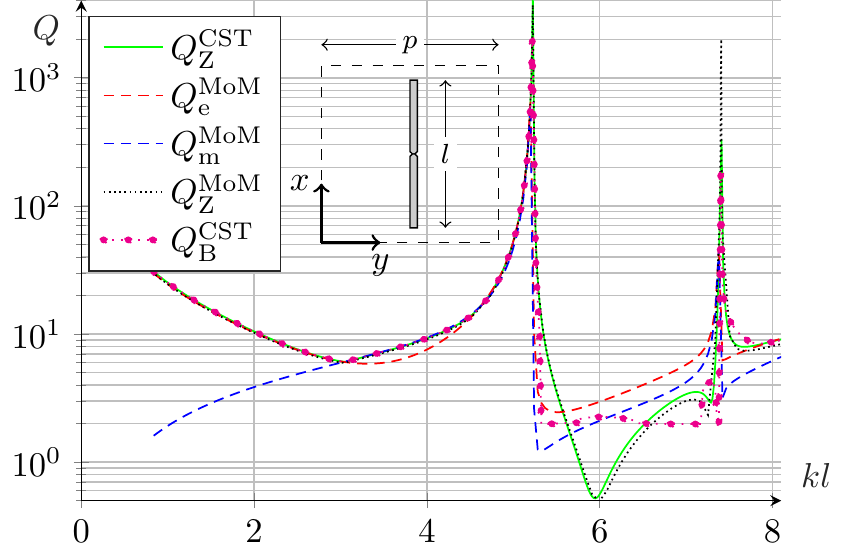}
	
	(b)
	\endgroup
	\end{minipage}
  \caption{Q-factor of a phased array with center-fed strip dipole elements (a) as a function of a scan angle $\theta_0$ in the E-plane ($\phi_0=0^\circ$) and the H-plane ($\phi_0=90^\circ$); the dipoles are aligned along $\phi_0=0^\circ$ direction, $l=0.45\lambda$, $w=0.02\lambda$, $p=0.5\lambda$, and (b) as a function of the electric size $kl$, with width $w=l/40$, period $p=1.2l$.
  }
  \label{fig:Q_dipole}
\end{figure}

The dependence of the Q-factor on the phasing angle (beam steering) is presented in Fig.~\ref{fig:Q_dipole}a for an array of center-fed dipoles, length $l=0.45\lambda$, width $w=0.02\lambda$, and periodicity $p=0.5\lambda$, see the inset in Fig.~\ref{fig:Q_dipole}b.
The dipoles are aligned along the $x$-axis; the scan angle $\theta_0$ is given in the E-plane ($\phi_0=0^\circ$) and the H-plane ($\phi_0=90^\circ$), where $\phi_0$ is counted from the positive $x$-semiaxis.
The here derived Q-factor in expression~\eqref{eq:Qe_Qm} is given by solid lines, and the Q-factors by our implementation of Kwon and Pozar method~\cite{Kwon+Pozar2014} are given by dashed lines.
Good agreement between the two methods is observed, and the small discrepancies are the outcome of different assumptions on current density in the methods. In~\cite{Kwon+Pozar2014} a single basis function with piece-wise sinusoidal distribution along the dipole's length, and uniform distribution along the dipole's width were used, and the current's direction was strictly along the dipole's length. To calculate the Q-factor with our expressions, we used an EFIE current solution for the model with 239 RWG basis functions and a voltage gap excitation.
Our model allows a non-uniform current distribution along the dipole's width as well as arbitrary current direction within the dipole's surface.

Fig.~\ref{fig:Q_dipole}b shows the Q-factor as a function of the electric length $kl$ for an array of center-fed strip dipoles. 
The dipole element has length $l$ and width $w=l/40$, the periodicity is $p=1.2l$ in both directions of the array plane.
Here a phasing angle $(\theta_0,\phi_0)=(0,0)$ is assumed, corresponding to a $z$-directed radiation. The voltage gap excitation is used.
The electric $Q_{\rm e}$ and magnetic $Q_{\rm m}$ Q-factors in~\eqref{eq:Qe_Qm} are shown by dashed red and blue curves respectively, where a model with 239 RWG functions was used.
The equivalent-tuned-bandwidth $Q_{\rm B}$, shown by magenta dotted line with markers, is calculated, for reflection threshold level $\Gamma_0=-10$ dB, from the voltage-gap input impedance, solved by CST solver.
The $Q_{\rm Z}$ from the input-impedance solutions by CST and by our in-house MoM code are shown with solid green and dotted black curves respectively.
We observe a fine agreement between all the curves at the frequencies below the first grating lobe, which occurs at $kl\simeq 5.2$.
Above this electrical length, all methods capture the grating-lobe peaks of the Q-factor with a good agreement, however differences are observed between the peaks.
The stored-energy Q-factor $\max (Q_{\rm e},Q_{\rm m})$ estimates the -10 dB bandwidth as expressed by $Q_{\rm B}$ better than $Q_{\rm Z}$ in the first half of that interval up till $kl\simeq 6.3$; above this value $Q_{\rm Z}$ appears to be closer to $Q_{\rm B}$.
Note that the array is sufficiently wideband between the first two grating lobes ($Q\leq 5$), and the Q-factor model might not be reliable there~\cite{Yaghjian+Best2005}.
At the grating lobe frequencies we observe that there are singularities of Q-factor. These singularities are related to the periodic resonance of the array.
At low frequencies, we see that the stored electric energy dominates, which is typical for dipole-type antennas.
The minimum of Q is located at $kl\simeq 3$, where the electric and the magnetic stored energies are equal. This electrical length corresponds to $l\approx 0.48 \lambda$, where $\lambda$ is the wavelength. This agrees with the length of a single dipole required to achieve its first self-resonance~\cite{Balanis1997}.

\subsection{Rectangular loop array}

\begin{figure}[htbp]
  \vspace{5pt}
  \centering 
  \begin{minipage}{0.5\textwidth}
  \centering 
 \begingroup
	\includegraphics[]{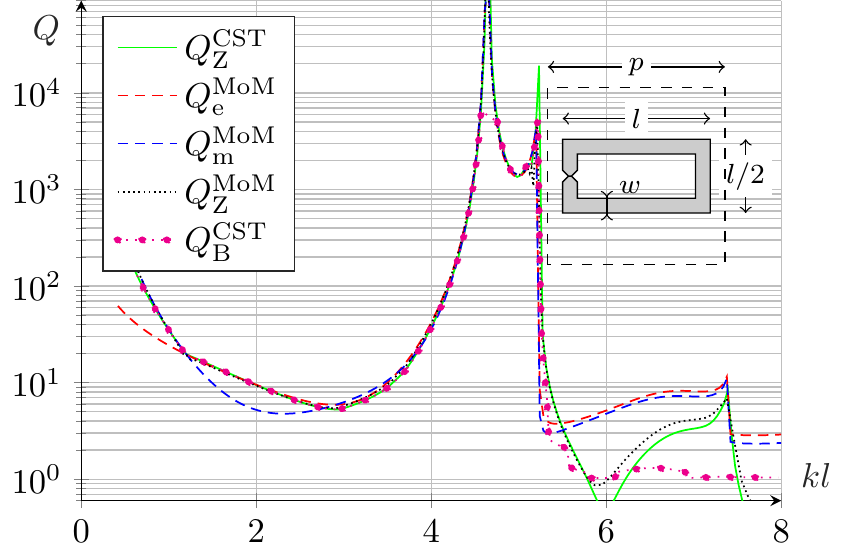}
  \endgroup
  
  (a)
  
   \end{minipage}
  \begin{minipage}{0.3\textwidth}
  \centering
  \includegraphics[scale=0.45]{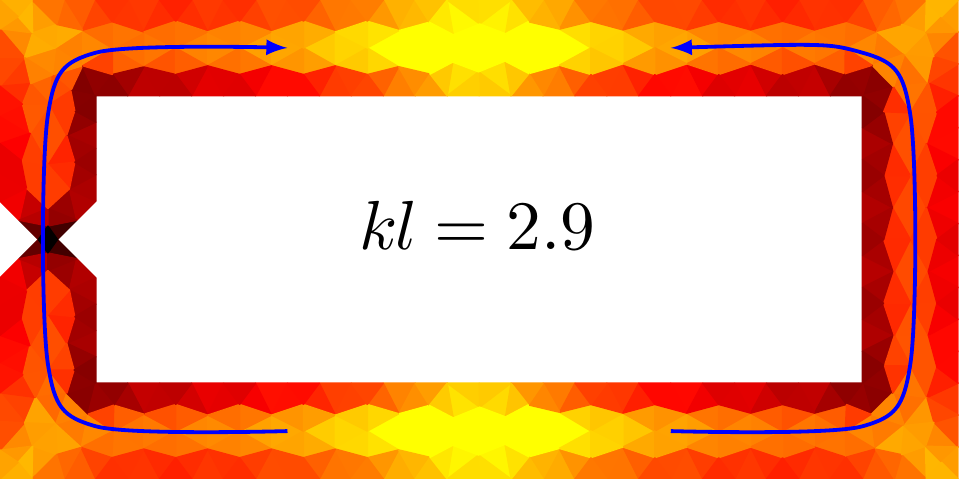}
  
  \vspace{5pt}
  
  \includegraphics[scale=0.45]{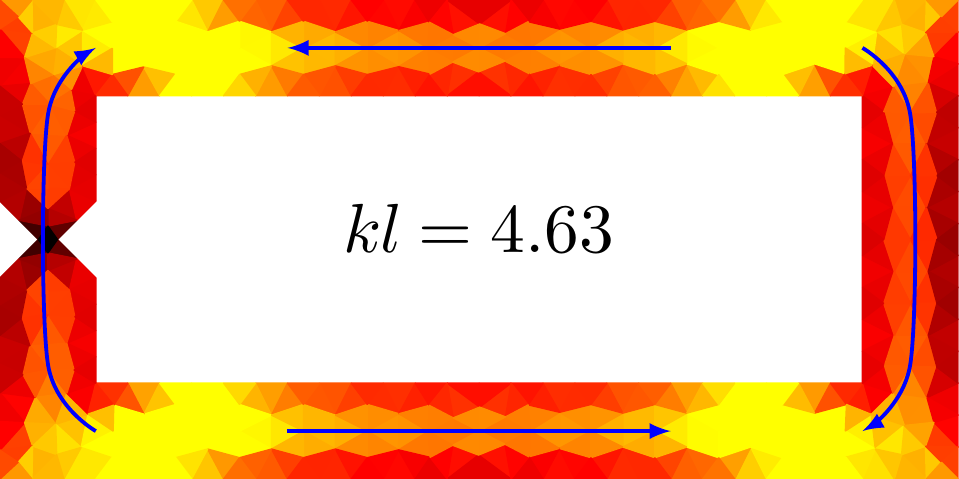}
  
  
  (b)
  \end{minipage}
    \begin{minipage}{0.07\textwidth}
  \centering
	\includegraphics[scale=0.8]{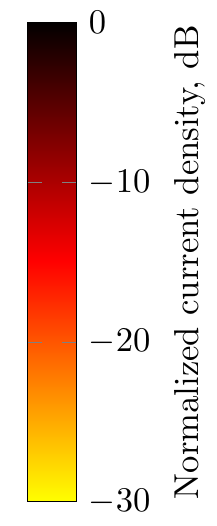}
    \vspace{10pt}
  \end{minipage}
  \caption{(a) Q-factor of a phased array with rectangular loop elements as a function of the electric size $kl$, with $w=l/10$, $p=1.2l$. (b) Current density distribution for $kl=2.9$ and (c) $kl=4.63$. The colormap varies from low current density magnitude (pale yellow) to high current density magnitude (dark red), the blue arrows show the direction of the current.}
  \label{fig:sloop}
\end{figure}

The next example is a rectangular loop (also called strip folded dipole) array.
Fig.~\ref{fig:sloop} depicts the Q-factors of such array with the strip width $w=l/10$ and periodicity $p=1.2l$.
Each loop is fed at the middle of one of its shorter sides, and a phasing angle $(\theta_0,\phi_0)=(0,0)$ is considered. 
The loop model for the in-house MoM code consists of 798 RWG basis functions. The colours of the Q-factor curves are the same as in Fig.~\ref{fig:Q_dipole}b in this and all the following examples.

A good agreement between the Q-factors is observed for frequencies below the grating lobe at $kl\simeq 5.2$.
The magnetic stored energy dominates at the low frequencies, which is common for the loop-type antennas.
The minimal value of the Q-factor below the grating lobe occurs at $kl\simeq 2.9$, where the magnetic and the electric stored energies are equal.
For this electric length, the current density distribution is depicted in Fig.~\ref{fig:sloop}b (upper), where the normalized magnitude of current density is shown by a colourmap (from pale yellow for low magnitude to dark red for high magnitude), and the instantaneous direction of the current is represented by the blue arrows.
We see that the length of the loop is approximately equal to one wavelength, which indicates a self-resonance of the loop.
The loop at this electric length can be seen as two bent half-wavelength dipoles, fed in phase.
Such a configuration has a maximum of its radiation pattern in the broadside direction~\cite{Balanis1997}. Thus the surface waves in the array plane are low.

As a contrast, we observe an additional peak of the Q-factor at $kl\simeq 4.6$, below the grating lobe. For this electric length, the current density distribution is depicted in Fig.~\ref{fig:sloop}b (lower), and here two wavelengths fit in the loop.
The loop can be seen as two symmetric pairs of half-wavelength dipoles, with a phase shift $180^\circ$ between dipoles in each pair.
Such a configuration has a null of the radiation pattern in the broadside direction, and a maximum in the  direction along the array plane.
In this regime, the surface waves dominate over the radiated waves, which yields a high peak value of the Q-factor.

A remark should be made regarding the nature of the grating lobe peaks in this and all the other examples; numerically those peaks can be made arbitrarily large in amplitude by refinement of the frequency sample grid in the neighbourhood of those peaks.
This indicates a singular behavior of the peaks, and also explains the differences in the peak amplitudes in numerical results by different methods.

Above the grating lobe we note that $Q_{\rm B}$ is below the prediction of $\max (Q_{\rm e},Q_{\rm m})$ and $Q_{\rm Z}$ and that the array is rather wideband $Q_{\rm B}\simeq 1-2$ in this region.

\subsection{Capped dipole array}

\begin{figure}[htbp]
  \vspace{5pt}
  \centering
  \begingroup
	\includegraphics[]{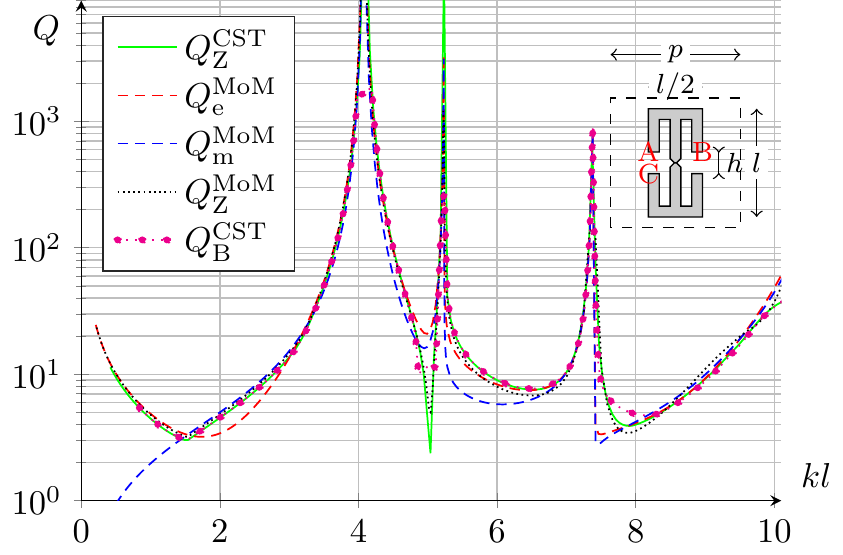}
  \endgroup
  ~
  \caption{The Q-factor of an array of capped dipoles as a function of the dipoles' electrical size $kl$, with strip width $w=l/10$, quadratic unit cell with $p=1.2l$.}
  \label{fig:Q_capped_dipole}
\end{figure}

The Q-factor comparison for an array of capped dipoles is presented in Fig.~\ref{fig:Q_capped_dipole} as a function of the electric size $kl$ of array elements.
The element parameters, as illustrated by an inset, are $p=1.2l$, and $h=l/5$, and the strip width $w=l/10$.
To compute the Q-factor, we used an element model with 931 RWG elements.
The curves agree well in the whole range of $kl$ below the first grating lobe, excluding the region around $kl\simeq 5$.
At this electrical size, the approximate length of the dipole's section between the ends A and B (see the inset in Fig.~\ref{fig:Q_capped_dipole}) is one wavelength $\lambda$.
The length of the section between points A and C (or B and C) is $\sim 1.5\lambda$.
Thus, a multiple resonance of the structure occurs, which induces a dip of the impedance Q-factor $Q_{\rm Z}$.
This effect is discussed for a finite case in~\cite{Gustafsson+Nordebo2006,Stuart+etal2007,Gustafsson+Jonsson2015}, where there is a deviation between $Q_{\rm B}$ and $Q_{\rm Z}$.
We observe a good agreement of all the methods as well above the first grating lobe, excluding a region around $kl\simeq 7.8$, where there is a slight difference between all the methods. 
As a contrast, recall Fig.~\ref{fig:Q_dipole}b, where the methods give different results in the range of $kl$ above the grating lobe.

\subsection{Array of mutually coupled dipoles}
\begin{figure}[htbp]
  \centering
  \centering
  \begingroup
	\includegraphics[]{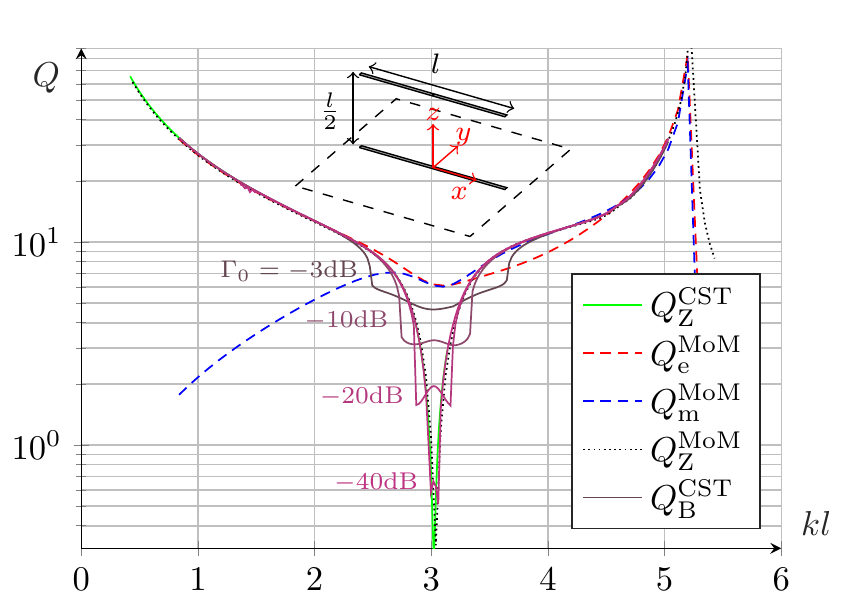}
  \endgroup

  \caption{The Q-factor of an array of mutually coupled dipoles as a function of the electric size $kl$, with width $w=l/40$, period $p=1.2l$. The tuned-bandwidth Q-factor $Q_{\rm B}$ is given for different reflection coefficient thresholds $\Gamma_0=-\{3,10,20,40\}$dB.}
  \label{fig:Q_yagi_uda}
\end{figure}

The Q-factors for an array of mutually coupled dipoles are shown in Fig.~\ref{fig:Q_yagi_uda}.
The unit cell consists of two dipole elements, a parasitic dipole in the plane $z=0$ and a driven element in the plane $z=l/2$.
The driven element is fed in its center.
Each dipole is comprised of 239 RWG elements and the lengths and widths of the dipoles are the same.
All the curves agree well in the whole range of $kl$ except of the interval $kl=[2.5, 3.5]$.
There, due to the multiple resonance of the structure, the $Q_{\rm Z}$ has a sharp dip~\cite{Gustafsson+Nordebo2006},~\cite{Stuart+etal2007},~\cite{Gustafsson+Jonsson2015}.
We observe that the equivalent tuned impedance Q-factor, with increase of the reflection coefficient threshold $\Gamma_0$ converges towards $Q_{\rm Z}$.
Indeed, the lower $\Gamma_0$ is, the more it highlights the single resonance caused by the matching element.
We observe that for this particular array, the proposed stored-energy Q-factor predicts bandwidth better for less severe thresholds $\Gamma_0\geq -10$dB, while the impedance-based $Q_{\rm Z}$ is more accurate when $\Gamma_0< -10$dB.

\subsection{Conical spiral array}

\begin{figure}[htbp]
  \centering
  \begin{minipage}{.40\linewidth}
  \centering
  	\includegraphics[width=0.7\linewidth]{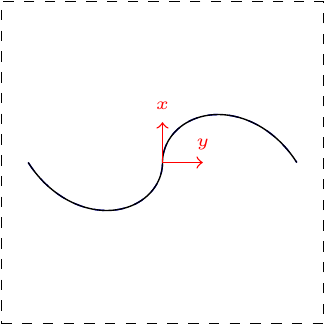}
    
    \includegraphics[width=0.7\linewidth]{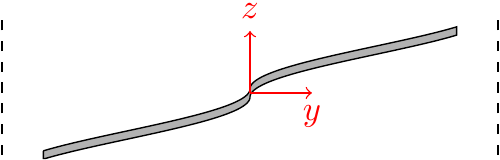}

    (a)
  \end{minipage}
    \caption{ The geometry of archimedian spiral elements for the parameter (a)~$N=0.25$ and (b)~$N=1.5$.}
\end{figure}

\begin{figure}[htbp]
  \centering
  \begingroup
	\includegraphics[]{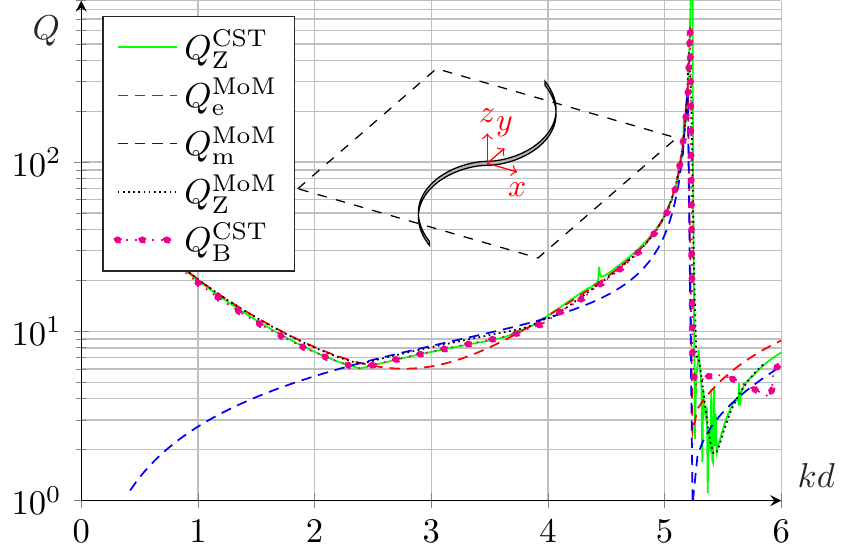}
  \endgroup

  \caption{The Q-factor of an array of center-fed conical Archimedian spirals with diameter $d$, strip width $w=d/20$, period $p=1.2d$, $N=0.25$.}
  \label{fig:Q_spiral}
\end{figure}

A conical Archimedian spiral array is here used as an example of a non-flat structure having electric current density with nonzero $z$-component.
The parametric equations, governing the Archimedian spiral's center line is
\begin{equation}
\begin{cases}
x &= t\cos(2\pi N t/d) \\
y &= t\sin(2\pi N t/d) \\
z &= \alpha t  \\
t &\in [-d,d].
\end{cases}
\end{equation}
Here we let $\alpha=0.3$, the strip width $w=d/20$ and the period $p=1.2d$.
The spiral geometry and the computed Q-factors for $N=0.25$ are depicted in Fig.~\ref{fig:Q_spiral}.
We observe a fine agreement between the methods below the grating lobe.
The minimum of Q-factor is at electric size $kd=2.3$, which corresponds to the spiral's length equal approximately to half wavelength.
Above the grating lobe, all the methods give slightly different Q-factor values.

\subsection{Discussion}
In all presented examples the proposed stored-energy-based Q-factor agrees well overall at the frequencies below the first grating lobe.
When the Q-factor is low ($Q\leq 5$), there exist  cases with disagreement between the proposed Q-factor and the fractional bandwidth, see region around $kl\simeq 3$ in Fig.~\ref{fig:Q_yagi_uda}.
In such regions, a sharp dip of Q-factor occurs, which typically related to multi-resonance behaviour~\cite{Gustafsson+Nordebo2006,Stuart+etal2007,Gustafsson+Jonsson2015}.

\section{Conclusions}
\label{sec:conclusions}
In this paper, the Q-factor expression is derived for lossless two-dimensionally-periodic arrays in free space.
The proposed expression is given in terms of the electric currents only, and thus it accounts for an exact shape of an array element.
The stored energies and radiated power, that enter the Q-factor formula, are expressed in the similar manner to the EFIE, typically used in MoM solvers, and hence the numerical implementation requires marginal modifications of an existing MoM code.

The numerical investigation indicates, that for narrow-band ($Q>5$) arrays the expression accurately predicts tuned fractional bandwidth, and agrees with input-impedance-$Q$ proposed in~\cite{Yaghjian+Best2005}.
The proposed representation permits multiple propagating Floquet modes, and our simulations in this regime (\ie at frequencies above the first grating lobe) demonstrate reasonable accuracy of bandwidth prediction.
The accuracy of such prediction above the first gratinglobe appears to be better for dipole-type structures than for loop-type arrays, which is a phenomenon to consider in the future.
In this region antenna elements tend to become electrically large with multiple resonances and the Q-factor description ceases to be valid.

The main advantages of the proposed Q-factor representation are: (i) it takes into account the exact shape of an array element and its current distribution, (ii) mathematically it has the same form as Q-factor expressions of finite radiators in~\cite{Vandenbosch2010,Jonsson+Gustafsson2015}, which is suitable for establishing fundamental bounds as done for finite case in \eg\cite{Gustafsson+etal2012,Tayli+etal2018,Shi+etal2017}, (iii) it does not require a specified input port and thus applicable to larger class of problems, (iv) numerical computation of stored-energies and radiated-power kernels is not more difficult than computation of Z-matrix in MoM.

To conclude, the proposed expression is fully consistent with and generalizes the previous work~\cite{Kwon+Pozar2014}, by extending the set of geometries and constraints for the Q-factor to include arbitrary-shaped metal inclusions on a periodic grid with arbitrary number of propagating modes.

\section*{Acknowledgment}
The authors thank S.~Amaya Maldonado for his earlier work on the in-house periodic MoM solver.
The authors would like to acknowledge the support of the Swedish foundation for strategic research (SSF) under the grant `Complex analysis and convex optimization for electromagnetic design' and the support of Vinnova center under the ChaseOn/iAA project.

\appendices

\section{Vanishing of $\Im \int_{\delta\U_{\rm top}} \vec{E}\times\vec{H}^*\cdot \hat{\vec{n}}\diff S$}
\label{app:vanishingImExH}
To show that $\Im \int_{\delta\U_{\rm top}} \vec{E}\times\vec{H}^*\cdot \hat{\vec{n}}\diff S$ vanishes we write out explicitly the electric $\vec{E}$ and magnetic $\vec{H}$ fields in terms of currents.
The electric field~\eqref{eq:E_phi_A} contains the scalar potential and vector potential terms.
We start with evaluating the scalar-potential term at $\vr_2\in\delta\U_{\rm top}$ which is located at $z_2=d/2$.
We notice that the evanescent-modes terms $(m,n)\in \Z^2\setminus\mathcal{P}$ are decaying exponentially with growth of $d$ as $\vr_1\in \delta U_{\rm top}$, and the scalar-potential term is thus
\begin{equation}
\begin{split}
&\nabla_1\phi(\vec{r}_1) \\&= 
\frac{\iu}{\omega\epsilon}\int_\Om \nabla_1\Gp(\vec{r}_1,\vec{r}_2)\nabla_2\cdot\vec{J}(\vec{r}_2) \diff v_2 + \mathcal{O}(\eu^{-|k_{\rm z}|d/2}) \\&=
\frac{1}{2\omega\epsilon a b} \hspace{-0.5em} \underset{(m,n)\in \mathcal{P}}{\sum} \frac{-\iu\vec{k}_{mn}}{\Kzmn}\eu^{-\iu \vec{k}_{mn}\cdot \vec{r}_1} \int_\Om \eu^{\iu \vec{k}_{mn}\cdot \vec{r}_2}\nabla_2\cdot\vec{J}(\vec{r}_2) \diff v_2\\& + \mathcal{O}(\eu^{-|k_{\rm z}|d/2}).
\end{split}
\raisetag{\normalbaselineskip}
\end{equation}
Here, $k_{\rm z}=\min_{(m,n)\in \Z^2\setminus\mathcal{P}}|\Kzmn|$ is the smallest longitudinal wavenumber among evanescent modes, and $\vec{k}_{mn}=\Ktmn+\vhat{z} \Kzmn$.
Using the vector identity for a divergence of vector and scalar functions product, we obtain
\begin{equation}
\begin{split}
&\nabla_1\phi(\vec{r}_1) \\&= 
\frac{1}{2\omega\epsilon a b} \underset{(m,n)\in \mathcal{P}}{\sum} \frac{\iu\vec{k}_{mn}}{\Kzmn}\eu^{-\iu \vec{k}_{mn}\cdot \vec{r}_1} \int_\Om \nabla_2\eu^{\iu \vec{k}_{mn}\cdot \vec{r}_2}\cdot\vec{J}(\vec{r}_2) \diff v_2 \\
&+ \mathcal{O}(\eu^{-|k_{\rm z}|d/2}) \\ &=
\frac{\eta}{2k a b} \underset{(m,n)\in \mathcal{P}}{\sum} \frac{-\vec{k}_{mn}}{\Kzmn}\eu^{-\iu \vec{k}_{mn}\cdot \vec{r}_1} \vec{k}_{mn}\cdot \mathbfcal{J}_{mn} + \mathcal{O}(\eu^{-|k_{\rm z}|d/2}). 
\label{eq:grad_phi}
\end{split}
\end{equation}
where $\mathbfcal{J}_{mn}=\int_\Om \eu^{\iu \vec{k}_{mn}\cdot \vec{r}_2}\vec{J}(\vec{r}_2) \diff v_2$.
We express the vector potential~\eqref{eq:A_G} component in a similar form
\begin{equation}
\iu \omega \vec{A} = 
\frac{\eta k}{2a b} \underset{(m,n)\in \mathcal{P}}{\sum} \frac{1}{\Kzmn}\eu^{-\iu \vec{k}_{mn}\cdot \vec{r}_1} \mathbfcal{J}_{mn} + \mathcal{O}(\eu^{-|k_{\rm z}|d/2}).
\label{eq:jwA}
\end{equation}
The magnetic field vector is treated in a similar manner, employing the vector identity for a curl of vector and scalar functions product
\begin{equation}
\begin{split}
\vec{H}^*(\vec{r}_1) &= \frac{1}{\mu}\nabla_1\times\vec{A}^*(\vec{r}_1) =
\int_\Om \nabla_1\Gp^*(\vec{r}_1,\vec{r}_2)\times\vec{J}^*(\vec{r}_2) \diff v_2 \\&+ \mathcal{O}(\eu^{-|k_{\rm z}|d/2})\\ &=
\frac{-1}{2a b} \underset{(p,q)\in \mathcal{P}}{\sum} \frac{1}{k_{{\rm z}pq}}\eu^{\iu \vec{k}_{pq}\cdot \vec{r}_1} \vec{k}_{pq}\times \mathbfcal{J}^*_{pq} + \mathcal{O}(\eu^{-|k_{\rm z}|d/2}).
\label{eq:H}
\end{split}
\end{equation}

The orthogonality of the Floquet modes, instrumental in the next step, follows from the integral over the top surface $\delta\U_{\rm top}$ of the unit cell
\begin{equation}
\begin{split}
\int_{\delta\U_{\rm top}}& \eu^{-\iu(\Ktmn-\vec{k}_{tpq})\cdot\vec{\rho}} \diff s \\&=
\int_{x=0}^a \int_{y=0}^b \eu^{-\iu(\Ktmn-\vec{k}_{tpq})\cdot(x\hat{\vec{x}} + y\hat{\vec{y}})} \diff x \diff y =
S\delta_{mp}\delta_{nq}.
\label{eq:orthogonality}
\end{split}
\end{equation}
Note that this result is independent of the $z$-coordinate and is valid for any surface, which is a crossection of $\U$, parallel to the $xy$-plane.

Orthogonality of the modes allows us to reduce the order of summation in the following integral, where the electric and magnetic fields are substituted using expressions~\eqref{eq:grad_phi},~\eqref{eq:jwA},~\eqref{eq:H}
\begin{equation}
\begin{split}
&\int_{\delta\U_{\rm top}} \vec{E}\times\vec{H}^*\cdot \hat{\vec{n}}\diff S  \\&=
\underset{(m,n)\in \mathcal{P}}{\sum} \underset{(p,q)\in \mathcal{P}}{\sum} 
\int_{\delta\U_{\rm top}}
\eu^{-\iu(\vec{k}_{mn} -\vec{k}_{pq})\cdot \vec{r}_1} \diff S_1 \hat{\vec{z}} \\ &\cdot
\frac{\eta\left[ k^2\mathbfcal{J}_{mn} - \vec{k}_{mn} (\vec{k}_{mn}\cdot \mathbfcal{J}_{mn}) \right]
\times
\left[ \vec{k}_{pq}\times\mathbfcal{J}_{pq} \right]
}{4k(ab)^2\Kzmn k_{{\rm z}pq}} \\ &+ \mathcal{O}(\eu^{-|k_{\rm z}|d/2}) \\ &=
\hspace{-0.6em}\underset{(m,n)\in \mathcal{P}}{\sum}\hspace{-0.6em}
\hat{\vec{z}}\cdot
\frac{\eta\left[ k^2\mathbfcal{J}_{mn} - \vec{k}_{mn} (\vec{k}_{mn}\cdot \mathbfcal{J}_{mn}) \right]
\times
\left[ \vec{k}_{mn}\times\mathbfcal{J}_{mn} \right]
}{4kab\Kzmn^2} \\&+ \mathcal{O}(\eu^{-|k_{\rm z}|d/2}) \\ &= 
\underset{(m,n)\in \mathcal{P}}{\sum}
\frac{\eta
}{4kab\Kzmn}\left[ k^2|\mathbfcal{J}_{mn}|^2 - |\vec{k}_{mn}\cdot\mathbfcal{J}_{mn}|^2\right] \\&+ \mathcal{O}(\eu^{-|k_{\rm z}|d/2}),
\end{split}
\raisetag{\normalbaselineskip}
\end{equation}
where the vector triple product identity was used in the last transition.
It is evident that the expression is purely real, and thus $\Im \int_{\delta\U_{\rm top}} \vec{E}\times\vec{H}^*\cdot \hat{\vec{n}}\diff S = 0$.
Similarly $\Im \int_{\delta\U_{\rm bottom}} \vec{E}\times\vec{H}^*\cdot \hat{\vec{n}}\diff S = 0$ for the bottom surface of $\U_d$ as $d\to\infty$

\section{Vanishing of $\Re \int_\U\{ \nabla\cdot(\phi\nabla\phi^*)\}\diff v$}\label{app:term1_vanishing}
We apply the divergence theorem to the divergence term:
\begin{equation}
\Re \int_\U\{ \nabla\cdot(\phi\nabla\phi^*)\}\diff v=
\Re \int_{\delta\U}\{ \hat{\vec{n}}\cdot(\phi\nabla\phi^*)\}\diff s.
\label{eq:div_phi_nabla_phi}
\end{equation}
We note that $\{(\nabla\phi(\vec{r}), \phi(\vec{r})\}\eu^{\iu \vec{k}_{\rm t00} \cdot \vec{r}})$ are periodic with lattice vectors $a\vhat{x}$ and $b\vhat{y}$, and thus $\phi \nabla \phi^*$ is periodic.
Hence the integration over $\delta \U_{\rm sides}$ at the right-hand side of~\eqref{eq:div_phi_nabla_phi} vanishes.

At $\delta\U_{\rm top}$, the normal component of the gradient factor~\eqref{eq:grad_phi} is
\begin{equation}
\begin{split}
\hat{\vec{z}}\cdot\nabla \phi^*(\vec{r}_1) &= \frac{-\eta}{2k a b} \underset{(p,q)\in \mathcal{P}}{\sum} \eu^{\iu \vec{k}_{pq}\cdot \vec{r}_1} \vec{k}_{pq}\cdot \mathbfcal{J}^*_{pq} + \mathcal{O}(\eu^{-|k_{\rm z}|d/2}). 
\end{split}
\end{equation}
Similarly the scalar potential~\eqref{eq:phi_J} is
\begin{equation}
\phi(\vec{r}_1) = \frac{-\eta}{2k a b}\hspace{-0.5em} \underset{(m,n)\in \mathcal{P}}{\sum} \hspace{-0.3em} \frac{\iu}{\Kzmn}\eu^{-\iu \vec{k}_{mn}\cdot \vec{r}_1} \vec{k}_{mn}\cdot \mathbfcal{J}_{mn} + \mathcal{O}(\eu^{-|k_{\rm z}|d/2}). 
\end{equation}

Combination of the last two equations with the orthogonality of modes~\eqref{eq:orthogonality} gives
\begin{equation}
\begin{split}
&\Re \int_{\delta\U_{\rm top}}\{ \hat{\vec{z}}\cdot(\phi\nabla\phi^*)\}\diff S \\ &=
\Re \frac{\eta^2}{4k^2ab} \underset{(m,n)\in \mathcal{P}}{\sum }\frac{\iu}{\Kzmn} |\vec{k}_{mn}\cdot \mathbfcal{J}_{mn}|^2 + \mathcal{O}(\eu^{-|k_{\rm z}|d/2}) \\&= \mathcal{O}(\eu^{-|k_{\rm z}|d/2}),
\end{split}
\raisetag{\normalbaselineskip}
\end{equation}
which vanish as $d\to\infty$.
Similarly the integral over $\U_{\rm bottom}$ vanishes, and thus the integral over the whole $\delta \U$ vanishes.

\section{Order of integration and summation}
\label{app:term2}

In the first line of expression~\eqref{eq:We2}, the integration and summation order is as follows
\begin{equation}\label{II}
\begin{split}
    &\int_{\U \times \Om \times \Om} 
    \underset{(m,n,p,q)\in \Z^4\setminus\mathcal{P}^2}{\sum}
    \G_{mn}(\vr,\vr_1) \G_{pq}^*(\vr,\vr_2) \\
    &\hspace{11em}\vec{J}(\vr_1)\cdot \vec{J}^*(\vr_2) 
    \diff v_1 \diff v_2 \diff v,
    \end{split}
\end{equation}
where $G_{mn}$ is a spectral term of the periodic Green's function~\eqref{eq:Greens_function}.
For efficient and convenient evaluation of such expression, it is desirable to move the integration over the unit cell $\U$ under both the integral over $\Om \times \Om$ and the infinite sum.
Performing unit-cell-volume integration term-by-term significantly simplifies the expression, however interchange of the orders of operations (integration and infinite summation) may in general affect the result, and thus must be justified. 
We here consider \eqref{II} in the case where the frequency is in the open set defined by removing the grating-lobe frequencies, that is $|k_{zmn}|>0$. At the grating-lobe frequencies it is well known that the Floquet-mode representation of the Green's function diverges.

The first step is to show that we are able to interchange the infinite summation and integration over $\U \times \Om \times \Om$.
We start with showing that the interchange is valid for the integration over $[\U \times \Om \times \Om]_\delta = \{(\vr,\vr_1,\vr_2): |z-z_1|\geq\delta, |z - z_2|\geq\delta\}$.
To connect with the original problem, we note that $\lim_{\delta\to 0}\int_{[\U \times \Om \times \Om]_\delta} = \int_{\U \times \Om \times \Om}$.
As the summations over $(m,n)$ and $(p,q)$ are independent of each other, it suffices to show $\int_{\U\times\Om^2} \underset{(m,n)\in \Z^2\setminus\mathcal{P}}{\sum}  \G^*_{pq} \G_{mn} \vec{J}_1\cdot \vec{J}^*_2 \diff v_1 v_2 v = \underset{(m,n)\in \Z^2\setminus\mathcal{P}}{\sum} \int_{\U\times\Om^2} \G^*_{pq} \G_{mn} \vec{J}_1\cdot \vec{J}^*_2 \diff v_1 v_2 v$ for fixed $(p,q)$.
Thus, we consider closer the convergence of the sum in $(m,n)$ by focusing in each term on the factors, dependent on $m, n$.

Define the sequence of functions
\begin{equation}
    f_N (\vr) =  \underset{\vec{n}\in (\Z^2\setminus \mathcal{P})\cap |\vec{n}|\leq N}{\sum}\frac{1}{\Kzmn}\eu^{-\iu \Ktmn\cdot\vec{\rho}_0-|\Kzmn||z_0|}.
\end{equation}
Here, we introduced the multi-index  $\vec{n}=(m,n)$, $|\vec{n}|^2=m^2+n^2$.
Note that there always exists  $n_0$ such that the longitudinal wavenumber can be estimated as $|\Kzmn|\geq \alpha|\vec{n}|$ with some $\alpha>0$ whenever $|\vec{n}|\geq n_0$.
For the purposes of the proof, it is sufficient to consider  $N\geq n_0$ such that all $\vec{n}$ with $|\vec{n}|\geq n_0$ are not in $\mathcal{P}$.

We show that $\{f_N\}$ is uniformly convergent on $F=\{\vr_0\in\R^3:x_0\in[0,a],y_0\in[0,b],z_0\in\R,|z_0|\geq\delta\}$ by estimating
\begin{equation}
\begin{split}
    |f_N - f_M| &= \left| \underset{\vec{n}: M'<|\vec{n}|\leq N'}{\sum}
    \frac{1}{\Kzmn}\eu^{-\iu \Ktmn\cdot\vec{\rho}_0-|\Kzmn||z_0|}\right| \\ &
    \leq
    \underset{\vec{n}}{\sum}
    \frac{\eu^{-\alpha |\vec{n}||z_0|}}{\alpha |\vec{n}|}
    \leq
    \frac{1}{\alpha}\int_0^{2\pi}\diff \theta \int_{M'-\sqrt{2}}^{N'}\frac{\eu^{-\alpha\delta r}}{r} r\diff r
    \\&\leq \frac{2\pi}{\alpha^2 \delta} \eu^{-\alpha \delta (M'-\sqrt{2})},
\end{split}
\raisetag{\normalbaselineskip}
\label{eq:unif_conv}
\end{equation}
where $M'=\min(M,N)\geq n_0$ and $N'=\max(M,N)$.
This result is independent of $\vec{\rho}_0$ and $z_0$, and can be made arbitrarily small by choosing sufficiently large $M'$.
Thus, by the Cauchy criterion for uniform convergence, see \eg\cite[Thm.~7.8]{Rudin1976}, $\{f_N\}$ converges uniformly on $F$.
The uniform convergence allows us to interchange the integration and summation $\int_A f_\infty \diff v=\lim_{N\to \infty}\int_A f_N\diff v$~\cite[Thm.~7.16]{Rudin1976}.
Applying this twice yields
\begin{equation}
\begin{split}
    &\int_{[\U \times \Om \times \Om]_\delta} 
    \underset{(m,n,p,q)\in \Z^4\setminus\mathcal{P}^2}{\sum}
    \G_{mn}(\vr,\vr_1) \G_{pq}^*(\vr,\vr_2) \\
    &\hspace{11em}\vec{J}(\vr_1)\cdot \vec{J}^*(\vr_2) 
    \diff v_1 \diff v_2 \diff v \\&= 
    \underset{(m,n,p,q)\in \Z^4\setminus\mathcal{P}^2}{\sum}
    \int_{[\U \times \Om \times \Om]_\delta} \G_{mn}(\vr,\vr_1) \G_{pq}^*(\vr,\vr_2) \\ &\hspace{11em}\vec{J}(\vr_1)\cdot \vec{J}^*(\vr_2) \diff v_1 \diff v_2 \diff v.
    \end{split}
\end{equation}

Next, we show that for each term in the sum above, the order of integration is interchangeable, that is
\begin{equation}
\begin{split}
    &\int_{[\U \times \Om \times \Om]_\delta} \G_{mn}(\vr,\vr_1) \G_{pq}^*(\vr,\vr_2) \\ &\hspace{11em}\vec{J}(\vr_1)\cdot \vec{J}^*(\vr_2) \diff v_1 \diff v_2 \diff v \\
    &=\int_{[\Om \times \Om\times\U]_\delta} \G_{mn}(\vr,\vr_1) \G_{pq}^*(\vr,\vr_2) \\ &\hspace{11em}\vec{J}(\vr_1)\cdot \vec{J}^*(\vr_2) \diff v \diff v_1 \diff v_2.
    \end{split}
    \label{eq:int_interchange}
\end{equation}
By Fubini's theorem~\cite[Thm. 2.16.4]{Friedman1982}, for any $h$ measurable function on $X\times Y$, if either $\int\int |h|\diff \mu \diff \nu < \infty$ or $\int\int |h|\diff \nu \diff \mu < \infty$, the result of an integral of $h$ over $X\times Y$ is independent of the order in which integrals are taken.

To see that it applies to our case, we first note that the function
\begin{equation}
\begin{split}
    K(\vr_1,\vr_2) = \int_{\U\cap\{|z-z_i|\geq \delta, i=1,2\} } |\G_{mn}(\vr,\vr_1) \G_{pq}^*(\vr,\vr_2)| \diff v
    \end{split}
\end{equation}
is bounded by a constant, independent of $\vr_1,\vr_2$.
Thus
\begin{equation}
\begin{split}
    \int_{[\Om \times \Om\times\U]_\delta} \hspace{-3em} |\G_{mn}(\vr,\vr_1) \G_{pq}^*(\vr,\vr_2) \vec{J}(\vr_1)\cdot \vec{J}^*(\vr_2)| \diff v \diff v_1 \diff v_2
    \end{split}
\end{equation}
is finite for absolutely integrable currents $\vec{J}\in L_1$.
Such currents are measurable since they are in $L_1$, and $G_{mn}G_{pq}^*$ is continuous and hence measurable on $[\Omega\times\Omega\times U]_\delta$.
Thus, $\G_{mn}(\vr,\vr_1) \G_{pq}^*(\vr,\vr_2) \vec{J}(\vr_1)\cdot \vec{J}^*(\vr_2)$ is measurable, since products of measurable functions are measurable~\cite[Thm 11.18]{Rudin1976}.
This satisfies the conditions of Fubini theorem, and hence~\eqref{eq:int_interchange} holds true.

By applying the convergence argument of~\eqref{eq:unif_conv}, we can interchange $\int_{[\Om \times \Om\times\U]_\delta}$ and summation.
Finally, by taking limits in $\delta$, we obtain
\begin{equation}\label{III}
\begin{split}
    &\int_{\U\times\Om \times \Om}
    \underset{(m,n,p,q)\in \Z^4\setminus\mathcal{P}^2}{\sum}
    \G_{mn}(\vr,\vr_1) \G_{pq}^*(\vr,\vr_2) \\
    &\hspace{11em}\vec{J}(\vr_1)\cdot \vec{J}^*(\vr_2)
    \diff v_1 \diff v_2 \diff v \\&= 
    \int_{\Om \times \Om\times \U}
    \underset{(m,n,p,q)\in \Z^4\setminus\mathcal{P}^2}{\sum}
    \G_{mn}(\vr,\vr_1) \G_{pq}^*(\vr,\vr_2) \\
    &\hspace{11em}\vec{J}(\vr_1)\cdot \vec{J}^*(\vr_2)
    \diff v \diff v_1 \diff v_2.
    \end{split}
\end{equation}

The next step is to reduce the
kernel $g$, \ie the integral over $U$ and the sum over $G_{mn}G_{pq}^*$ in ~\eqref{III} as compactly as possible. 
We observe that $g$ in \eqref{eq:low_case_g} is equivalent with
\begin{equation}
\begin{split}
    g(\vec{r}_1,\vec{r}_2)=&
    \int_\U 
    \underset{(m,n,p,q)\in \Z^4\setminus\mathcal{P}^2}{\sum}
    \G_{mn}\G_{pq}^* \diff v.
    \label{eq:sum_sum_g}
\end{split}
\end{equation}
Again, the interchange of the orders of operations (integration and infinite summation) may in general affect the result, and thus must be justified.
We first show that the interchange is valid for integration over $\U_\delta=\{\vr=(x,y,z)\in\R^3:x\in[0,a],y\in[0,b],z\in\R\setminus(B_1\cup B_2)\}$, where $B_i=\{z\in\R:|z-z_i|>\delta\}$, and subsequently demonstrate that the assertion holds also in the limit $\delta\to 0$.

From the uniform convergence argument~\eqref{eq:unif_conv} on $\U_\delta$ it follows that
\begin{equation}
\begin{split}
    &\int_{\U_\delta} 
    \underset{(m,n,p,q)\in \Z^4\setminus\mathcal{P}^2}{\sum}
    \G_{mn}(\vr,\vr_1) \G_{pq}^*(\vr,\vr_2)
    \diff v \\&= 
    \underset{(m,n,p,q)\in \Z^4\setminus\mathcal{P}^2}{\sum}
    \int_{\U_\delta} \G_{mn}(\vr,\vr_1) \G_{pq}^*(\vr,\vr_2) \diff v.
    \end{split}
\end{equation}
We integrate over the cross-section, utilizing the orthogonality of modes~\eqref{eq:orthogonality} to reduce the sum
\begin{equation}
\begin{split}
    &g(\vr_1,\vr_2) =\\&
    \lim_{\delta \to 0} 
    \underset{\vec{n}\in \Z^2\setminus\mathcal{P}}{\sum}
    \int_{\R\setminus B_1\cup B_2}\hspace{-1.5em}\diff z
    \frac{\eu^{\iu \Ktmn \cdot (\vec{\rho}_1-\vec{\rho}_2)}}{|\Kzmn|^2}
    \eu^{-|\Kzmn|(|z-z_1|+|z-z_2|)}.
    \end{split}
\end{equation}
It remains to show that the limit and the infinite sum are interchangeable. 
Define the sequence of functions
\begin{equation}
\begin{split}
    &h_N(\delta)=\\& \hspace{-1em}
    \underset{\vec{n}\in \Z^2\setminus\mathcal{P}
    \cap\{|\vec{n}|\leq N\}}{\sum}
     \int_{\R\setminus B_1\cup B_2}\hspace{-1.5em}\diff z
    \frac{\eu^{\iu \Ktmn \cdot (\vec{\rho}_1-\vec{\rho}_2)}}{|\Kzmn|^2}
    \eu^{-|\Kzmn|(|z-z_1|+|z-z_2|)}.
    \label{eq:seq_h}
\end{split}
\end{equation}
By \cite[Thm.~7.11]{Rudin1976}, $\lim_{\delta\to c} \lim_{N\to\infty}h_N(\delta) = \lim_{N\to\infty} \lim_{\delta\to c}h_N(\delta)$ holds if $\{h_N\}$ is uniformly convergent on a set $E$ and $\lim_{\delta\to c}h_N(\delta)$ exists, given $c$ is a limit point of $E$.

For the case $z_1\neq z_2$ take $E=(0,|z_1-z_2|/2)$, and $c=0$.
A straightforward integration yields
\begin{equation}
\begin{split}
    h_N(\delta)=
    \underset{\vec{n}\in \Z^2\setminus\mathcal{P}
    \cap\{|\vec{n}|\leq N\}}{\sum} &
    \frac{\eu^{\iu \Ktmn \cdot (\vec{\rho}_1-\vec{\rho}_2)}}{|\Kzmn|^2}
    \eu^{-|\Kzmn||z_1-z_2|} \\
    &\cdot
    \left\{ \frac{\eu^{-2|\Kzmn|\delta}}{|\Kzmn|} + 
    |z_1 - z_2| - 2\delta \right\}.
    \end{split}
\end{equation}
It is evident that the limit at $\delta \to 0$ exists:
\begin{equation}
\begin{split}
    \lim_{\delta\to 0} h_N(\delta) = 
    \underset{\vec{n}\in \Z^2\setminus\mathcal{P}
    \cap\{|\vec{n}|\leq N\}}{\sum}&
    \frac{\eu^{\iu \Ktmn \cdot (\vec{\rho}_1-\vec{\rho}_2)}}{|\Kzmn|^2}
    \eu^{-|\Kzmn||z_1-z_2|} \\&
    \left\{ \frac{1}{|\Kzmn|} + 
    |z_1 - z_2| \right\}.
    \end{split}
\end{equation}

To demonstrate the uniform convergence, we use Cauchy criterion
\begin{equation}
\begin{split}
    &|h_N(\delta)-h_M(\delta)| 
    \\&=
    \left|
    \underset{M'\leq|\vec{n}|\leq N'}{\sum}
    \frac{\eu^{\iu \Ktmn \cdot (\vec{\rho}_1-\vec{\rho}_2)}}{|\Kzmn|^2}
    \eu^{-|\Kzmn||z_1-z_2|} \right. \\&
    \left. \left\{ \frac{\eu^{-2|\Kzmn|\delta}}{|\Kzmn|} + 
    |z_1 - z_2| - 2\delta \right\} \right|
    \\ &\leq
    \underset{M'\leq|\vec{n}|\leq N'}{\sum}
    \frac{\eu^{-\alpha|\vec{n}||z_1-z_2|}}{\alpha^2|\vec{n}|^2}
    \left\{ \frac{1}{\alpha|\vec{n}|} + 
    |z_1 - z_2|\right\}
    \\ &\leq
    2\pi \int_{M'-\sqrt{2}}^{N'}
    \frac{\eu^{-\alpha|\vec{n}||z_1-z_2|}}{\alpha^2|\vec{n}|}
    \left\{ \frac{1}{\alpha|\vec{n}|}
    +|z_1 - z_2|\right\} \diff |\vec{n}|
     \\ &\leq
     \frac{2\pi}{\alpha^3} \left\{ \frac{1}{\alpha|z_1-z_2|}
    +1 \right\}
    \eu^{-\alpha |z_1-z_2| (M'-\sqrt{2})} .
\end{split}
\end{equation}
Here, $M'=\min(M,N)$ and $N'=\max(M,N)$.
This expression is independent of $\delta$ and can be made arbitrarily small by choosing sufficiently large $M'$, and thus $\{h_N(\delta)\}$ converges uniformly.
This, combined with existence of $\lim_{\delta\to 0} h_N(\delta)$ implies~\cite{Rudin1976} that  $\lim_{\delta\to 0}\lim_{N\to\infty}h_N(\delta)=\lim_{N\to\infty}\lim_{\delta\to 0}h_N(\delta)$ for $z_1\neq z_2$.

For the case $z_1=z_2$, we note that $B_1=B_2$ and take $E=(0,\infty)$.
The sequence~\eqref{eq:seq_h} after the direct integration is
\begin{equation}
\begin{split}
    h_N(\delta)=
    \underset{\vec{n}\in \Z^2\setminus\mathcal{P}
    \cap\{|\vec{n}|\leq N\}}{\sum}
    \frac{\eu^{\iu \Ktmn \cdot (\vec{\rho}_1-\vec{\rho}_2)}}{|\Kzmn|^3}
     \eu^{-2|\Kzmn|\delta}.
\end{split}
\end{equation}
We note the existence of the limit $\lim_{\delta\to 0} h_N(\delta)=h_N(0)$ for any $N$, and use Cauchy criterion for showing the uniform convergence
\begin{equation}
\begin{split}
    &|h_N(\delta)-h_M(\delta)| 
    \leq
    \underset{M'\leq|\vec{n}|\leq N'}{\sum}
    \frac{1}{|\Kzmn|^3}
    \leq
    \underset{M'\leq|\vec{n}|\leq N'}{\sum}
    \frac{1}{\alpha^3|\vec{n}|^3}
    \\ &\leq
    \frac{1}{\alpha^3}\int_0^{2\pi}\diff \theta\int_{M'-\sqrt{2}}^{N'}\frac{1}{|\vec{n}|^3}|\vec{n}|\diff |\vec{n}| \leq
    \frac{2\pi}{\alpha^3}\frac{1}{M'-\sqrt{2}}.
    \end{split}
\end{equation}
We note that this expression is independent of $\delta$ and can be made arbitrarily small by choosing appropriately large $M'$.
Thus the sequence is uniformly convergent, and the interchange of limits is valid for $z_1=z_2$ as well.
This completes the proof.

This allows us to perform integration in the kernel expression~\eqref{eq:sum_sum_g} term by term, which yields
\begin{equation}
\begin{split}
g(\vec{r}_1,\vec{r}_2)  
=\frac{1}{4 S} &
\underset{(m,n)\in \Z^2\setminus \mathcal{P}}{\sum} 
\frac{1}{|\Kzmn|^2} 
\eu^{\iu \Ktmn\cdot(\vec{\rho}_1-\vec{\rho}_2)} \\&
\eu^{-|\Kzmn||z_2-z_1|} \left( \frac{1}{|\Kzmn|} + |z_1-z_2|  \right).
\end{split}
\end{equation}




\end{document}